\RequirePackage{fix-cm}
\documentclass[smallextended]{svjour3}       % onecolumn (second format)
\smartqed  % flush right qed marks, e.g. at end of proof
\usepackage{graphicx}
\begin{document}

\title{The current orbit of Methone (S/2004 S 1)%\thanks{Grants or other notes
}


\author{Nelson Callegari Jr.         \and
        Adri\'an Rodr\'iguez and Dem\'etrio Tadeu Ceccatto %etc.
}


\institute{Nelson Callegari Jr. \at
             S\~{a}o Paulo State University (Unesp), Institute of Geosciences and Exact Sciences (IGCE). Avenue 24-A, 1515, Rio Claro, SP; Zip code 13506-900, Brazil.\\
              Tel.: +55-19-35269132\\
%              Fax: +123-45-678910\\
              \email{nelson.callegari@unesp.br}           %  \\
           \and
            Adri\'an Rodr\'iguez \at
              Observat\'orio do Valongo, Universidade Federal do Rio de Janeiro. Ladeira do Pedro Ant\^{o}nio 43, Rio de Janeiro, RJ;  Zip code 20080-090, Brazil.\\
               \email{adrian@astro.ufrj.br}
               \and
                Dem\'etrio Tadeu Ceccatto \at PhD student (Physics Program - IGCE/Unesp)  \\ \email{dtceccatto@gmail.com }
}

\date{Received: date / Accepted: date}

\maketitle

\begin{abstract}

The Cassini spacecraft discovered many close-in small satellites in Saturnian system, and several of them exhibit exotic orbital states due to interactions with Mimas and the oblateness of the planet. This work is devoted to Methone, which is currently involved in a 15:14 Mean-Motion Resonance with Mimas. We give an in deep study the current orbit of Methone by analyzing and identifying the short, resonant and long-term gravitational perturbations on its orbit. In addition, we perform numerical integrations of full equations of motion of ensembles of close-in small bodies orbiting the non-central field of Saturn. Spectral analyses of the orbits and interpretation of them in dynamical maps allow us to describe the orbit and the dynamics of Methone in view of resonant and long-term dynamics.

We show that the current geometric orbit of Methone is aligned with Mimas' due to a forced resonant component in eccentricity, leading to simultaneous oscillations of several critical angles of the expanded disturbing function. Thus, we explain the simultaneous oscillations of four critical arguments associated to the resonance.

The mapping of the Mimas-Methone resonance shows that the domains of the 15:14 Mimas-Methone resonance are dominated by regular motions associated to the Corotation resonance located at eccentricities lower than $\sim 0.015$ and osculating semi-major axis in the interval $194,660-194,730$ km. Methone is currently located deeply within this site.

\end{abstract}

\section{Introduction}

The detection of Methone was probably one of the first discoveries of the \emph{Cassini Mission} since it has been reported a few months after the insertion in the Saturnian system (Porco 2004). Methone is an ellipsoidal shaped, close-in small satellite with mean radius $\bar{R}=1.45\pm0.03$ km (Thomas and Helfenstein 2020). The first studies on orbital characterization of Methone were reported by Porco et al. (2005) and Spitale et al. (2006). In Porco et al. (2005), the dynamics and stability of test particles in the vicinity of Methone was also introduced. Spitale et al. (2006) proposed for the first time the proximity of Methone to a mean-motion orbital resonance with Mimas in order to explain many oscillatory components in the orbit of Methone. Spitale et al. (2006) suggested the angular combination $\psi_2\equiv15\lambda_{Me}-14\lambda_M-\varpi_{Me}$ as being the critical angle associated to the resonance, where $\lambda_{Me}$, $\lambda_M$ refer to the mean longitudes of Methone and Mimas, and $\varpi_{Me}$ to the longitude of pericenter of Methone. Jacobson et al. (2006a), followed by Jacobson et al. (2008), extended previous analyses on the Methone orbit and the 15:14 resonance. Like in Spitale et al. (2006), Jacobson et al. (2006a) also stated that the libration argument of the resonance is $\psi_2$. Hedman et al. (2009) considered the case of the arc of Methone in view of the 15:14 Mimas-Methone resonance. Hedman et al. (2009) also presented a puzzle characteristic of the orbit of Methone: they were the first to report that besides $\psi_2$, another angle, namely $\psi_1=15\lambda_{Me}-14\lambda_M-\varpi_{M}$ also oscillates around zero, where $\varpi_M$ is the longitude of pericenter of Mimas\footnote{We will define and denote the angles $\psi_2\equiv15\lambda_{Me}-14\lambda_M-\varpi_{Me}$ and $\psi_1\equiv15\lambda_{Me}-14\lambda_M-\varpi_{M}$ by the Lindblad and corotation angles, respectively. This nomenclature comes from planetary ring dynamics (e.g. Murray and Dermott 1999).}. However, no numerical simulation is given in Hedman et al. (2009) showing the \emph{simultaneous} libration of $\psi_1$ and $\psi_2$. El Moutamid et al. (2014) developed an analytical model given by an average hamiltonian appropriate for the study of the coupled equations of motion involving the terms of the disturbing function with both arguments, $\psi_1$ and $\psi_2$. They show in their figure 7 that $\psi_1$ is librating around zero, and in spite of the detailed analysis on the coupling between the resonant terms of $\psi_1$ and $\psi_2$, no discussion on the synchronicity of the angles $\psi_1$ and $\psi_2$ is given in the paper. A numerical simulation showing explicitly the concomitant oscillation of $\psi_1$ and $\psi_2$ is given for the first time in figures 1 and 12 in Mun\~oz-Guti\'errez and Giuliatti-Winter (2017), the latter covering the (long) time span of 10,000 yr.
%In other words, $\theta_i$ defines the reference direction along the line of the conjunctions of the resonant pair oscillates around due to resonance trapping.

At a first glance, the oscillations of two or more arguments for the same set of initial conditions and around the same center of libration may put in question what is classically known (e.g. Ferraz-Mello 1985, Peale 1999, Murray and Dermott 1999) in the theory of mean-motion resonant dynamics, because the line of the resonant conjunctions would not be well defined. However, let us consider an example with two critical angles $\theta_i$ and $\theta_j$ associated to the same mean-motion commensurability. In planetary dynamics, it is possible to write $\theta_j=\theta_i+\mu$, where $\mu$ represents a long-term component. Suppose that $\theta_i$ librates around a specific value, zero for instance. If $\mu$ circulates, $\theta_j$ will also circulate. Simultaneous oscillation of $\theta_j$ and $\theta_i$ \emph{is possible} when $\mu$ has an oscillatory behaviour. This scenario appears in 2:1 mean-motion planetary resonance, where the angle $\mu=\Delta\varpi_{j-i}$ is the relative variation of the longitudes of pericenters of the planets $i$ and $j$. Equilibria symmetric solutions of $\Delta\varpi_{j-i}$ can occur around 0 or $\pi$ in the case where strong forced components exist (e.g. Michtchenko and Ferraz-Mello 2001a). In this case, considering that the resonance is defined by the angle $\theta_i$, the libration of $\theta_j$ is just a cinematic consequence of the stationary variation of $\Delta\varpi_{j-i}$\footnote{On the contrary to the example given above, the dynamics of the resonance 15:14 Mimas-Methone is dictated within the domain of the restricted three-body problem (El Moutamid et al. 2014), since the mass of Mimas is much larger than Methone's.}.

In this work, we will explain in detail the mechanism of the multiple resonant oscillations in the orbit of Methone. For this task, first of all it is necessary a full investigation of the current orbit of Methone, as it is described in Section 3. The puzzle of the simultaneous oscillations of $\psi_1$ and $\psi_2$ is then solved in Section 4, where we show that this orbital configuration is only possible due to a forced component in the orbit of Methone raised by Mimas, leading to an alignment of the orbits of both satellites. We show that the 15:14 Mimas-Methone resonance contributes majority to this forced component. The alignment of the orbits occurs when we look to the \emph{geometric} orbit of Methone, because they are free from fluctuations due to the $J_2$ field of Saturn, as described in Section 2. On the contrary, when the osculating orbit of Methone is considered, the long-term alignment is just obfuscated by the large $J_2$ short-term perturbations. An additional result never reported before is that if we look in deep all multiplets associated to the 15:14 resonance, two other geometric critical angles oscillates together with $\psi_1$ and $\psi_2$.

In order to generalize the properties of the 15:14 Mimas-Methone resonance, in Section 5 we explore the resonance in the Methone's orbital elements phase space, where large sets of test small satellites are integrated and their orbits analyzed in appropriated planes of initial conditions. The results are the so-called dynamical maps, successfully applied in studies of resonant dynamics of other small satellites like Anthe (Callegari and Yokoyama 2020).

\begin{figure*}
\centering
 \includegraphics[width=6cm]{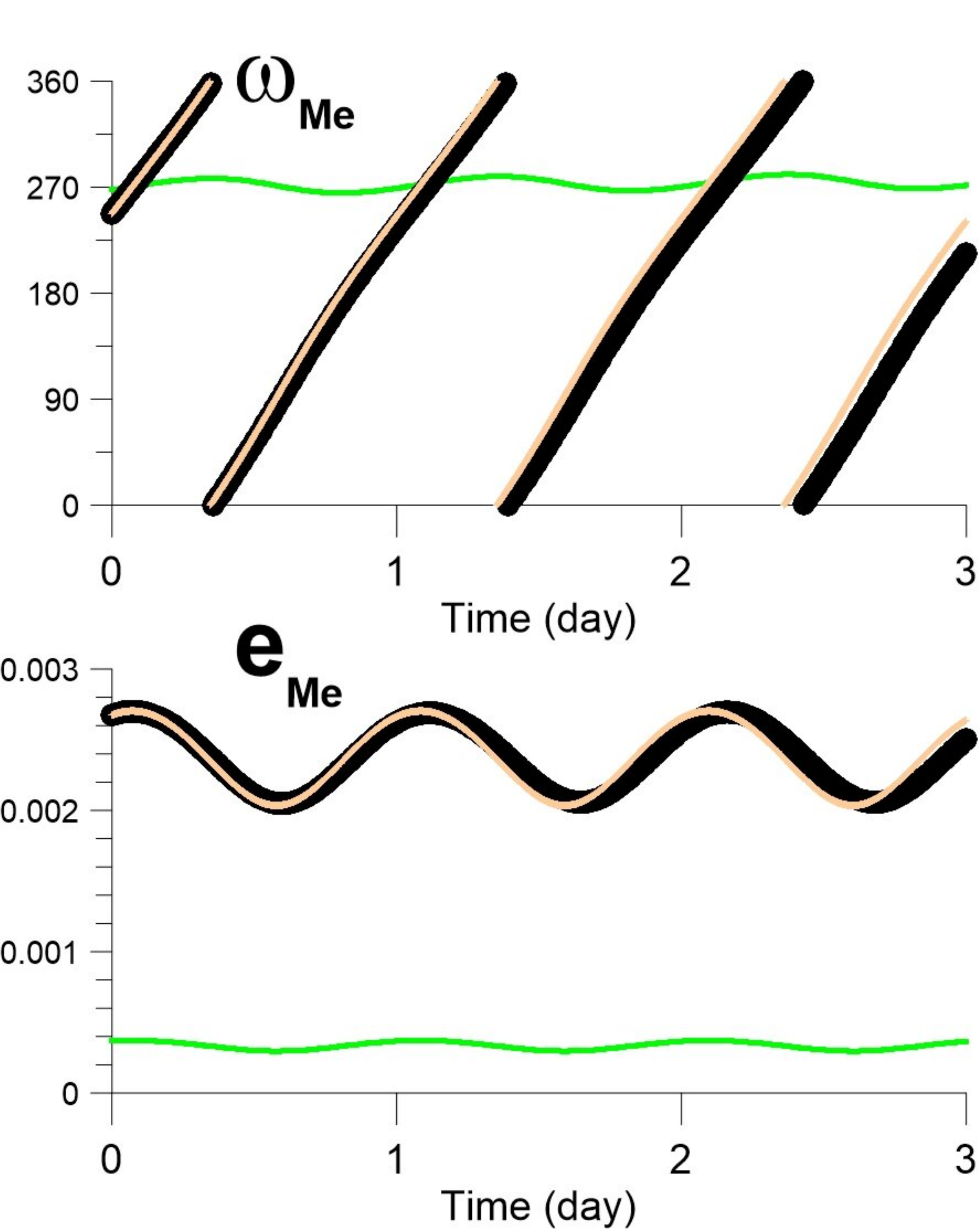}
  \caption{The argument of pericenter (top) and orbital eccentricity (bottom) of Methone. The black-extra-bold line is the osculating pericenter of Methone taken from http://ssd.jpl.nasa.gov/horizons.cgi with start time January 01, 2016. The data have been taken in the time span of 3 days with time step of 7 minutes. The pink curves have been generated from numerical scheme i), where only the $J_2$ perturbations has been taking into account. The green curves are the geometric pericenter (top) and eccentricity (bottom) of Methone calculated from the same simulation. }
  \label{<geo>}
\end{figure*}

%\newpage
\section{Methods}

In this work, we follow the methodology given in Callegari and Yokoyama (2010a, 2020) and numerically integrate the exact equations of motion of a system of $N$ satellites mutually disturbed orbiting under the actions of the main terms of the Saturn's potential expanded up to second order. The simulations can include up to seven satellites, namely, Methone plus Mimas, Enceladus, Tethys, Dione, Rhea and Titan. Two different models were adopted and compared to each other: i) Our own system of exact differential equation presented in Callegari and Yokoyama (2010a), where we may consider all satellites listed above and $J_2$, $J_4$ ($J_k$, $k$ even, are the dimensionless coefficient of the expanded potential); ii) Direct application of the Mercury package (Chambers 1999). In ii), the $J_6$ perturbations are added. In both cases i) and ii), we have applied the Everhart's code ``RA15'' (Everhart 1985), devoted to solve a system of ordinary differential equations with accuracy of fifteen order in local error of the numerical solution. The initial conditions and numerical parameters are listed in Appendix 1. The details \textbf{of} particular numerical simulations are always indicated in the caption of the corresponding figure (see Fig. \ref{<geo>} as an example).

Two additional efforts will be applied in order to analyze our set of numerical data: I) In some cases, the orbits will be directly compared to the \emph{osculating} orbital elements provided by the \emph{Horizons}' system of ephemerides (http://ssd.jpl.nasa.gov/horizons.cgi); II) From instantaneous positions and velocities of the satellites taken from numerical simulations, we calculate the so-called \emph{geometric} elements by applying the algorithm provided by Renner and Sicardy (2006). The geometric elements are in general calculated to better fit the orbit in the space when a close-in satellite revolves in a strong non-central field. In this case, the osculating argument of pericenter ($\omega$) advances periodically with the same frequency of the mean-motion. This component in $\omega$ is related to $J_2$ perturbations, a phenomenon known in classical works (e.g. Greenberg 1981). Black and pink curves in Fig. \ref{<geo>} have been generated, respectively, from \emph{Horizons} data and from a numerical simulation performed with the numerical scheme i), where only a satellite similar to Methone under the effect of $J_2$ has been included in the simulation. The good agreement between the curves shows that $J_2$ perturbation induces the short-term variation of $\omega$. The green curve is $\omega$ calculated with the numerical scheme II) from the vector states generated in the same simulation. Note that the circulatory regime gives place to small-amplitude fluctuations.

\begin{figure}
 \includegraphics[width=12cm]{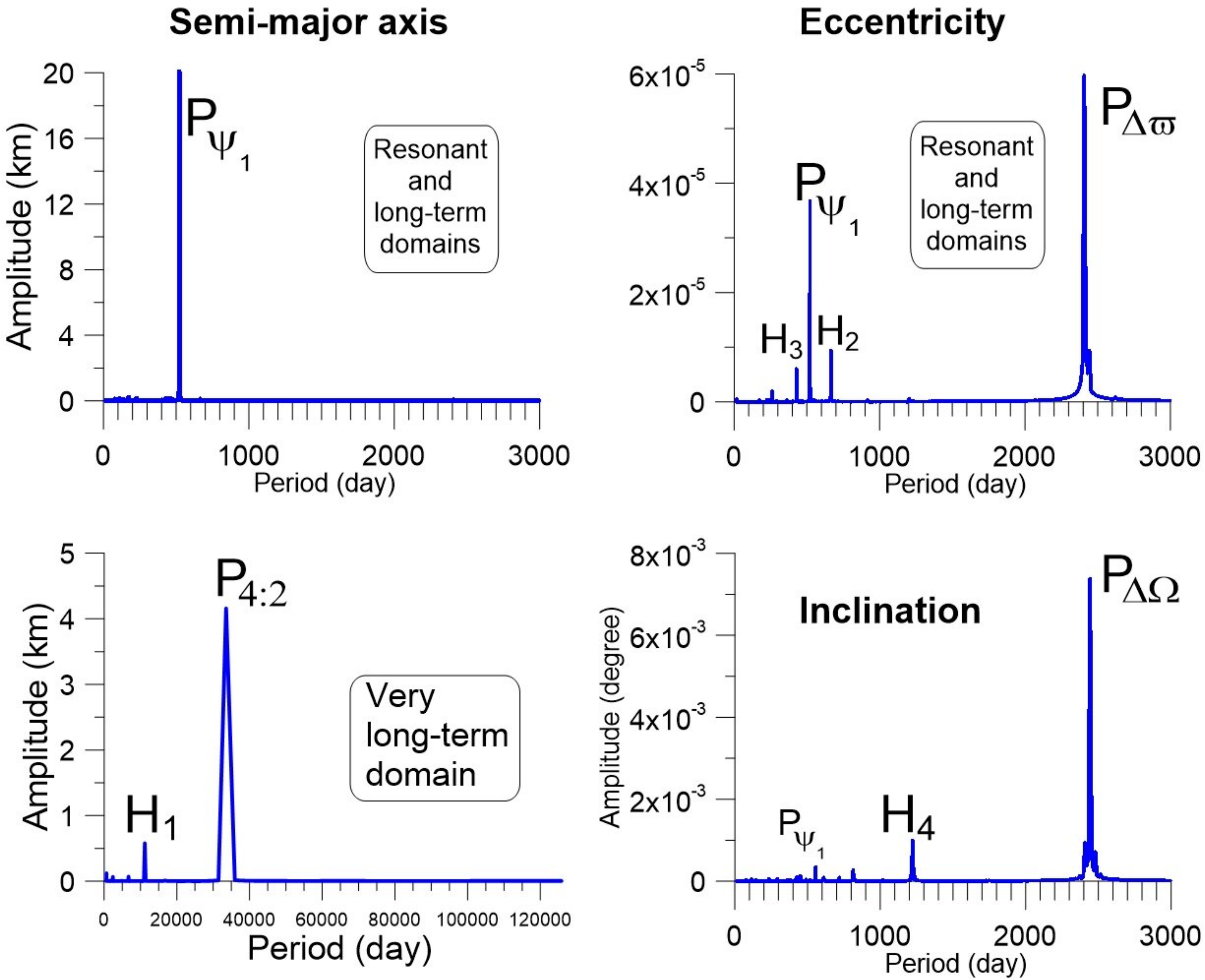}
  \caption{First column: spectrum of osculating semi-major axis of a satellite similar to Methone calculated with Mercury package (numerical scheme ii) described in Section 2). Initial conditions are given in Appendix 1. Amplitudes in semi-major axis are given in km. The total time of integration and sampling are 1720 year and 0.06 day, respectively. The main domains of the spectra discussed throughout the paper are indicated by balloons. The main peaks are also appointed and described in Table 1 and Sections 3.1-3.3. Note the different scales in y-axes of spectrum of the semi-major axis. Second column: spectra of osculating eccentricity (top) and inclination (bottom).}
  \label{<fft>}
\end{figure}

In addition, let us to mention here another method applied throughout this work: the orbits obtained numerically are analyzed in the frequency domain, so that, variables like the osculating semi-major axis, eccentricity and inclination are decomposed with the fast Fourier transform code provided by Press et al. (1996). Examples of spectra of a satellite with orbital properties similar to Methone are given in Fig. \ref{<fft>}. A great deal of significant peaks appear in distinct variables and are identified accordingly to different disturbing forces, as will be described in details in Section 3. Systematic scans of the spectra in large sets of initial conditions will allow to identify the dependence of the main amplitudes and corresponding frequencies. The results are presented in two different ways: the Individual Power Spectra (IPS) given in Fig. \ref{<ipsm>} in Section  5.1, and the dynamical maps like Fig. \ref{<dmm>} given in Section 5.2. Michtchenko and Ferraz-Mello (2001a,b) developed theses techniques, and we have introduced these approaches in resonant dynamics of mid-sized and inner small Saturnian satellites (Callegari and Yokoyama 2007, 2008, 2010a, 2020).

\section{The current orbit of Methone}

\begin{figure*}
 \includegraphics[width=12cm]{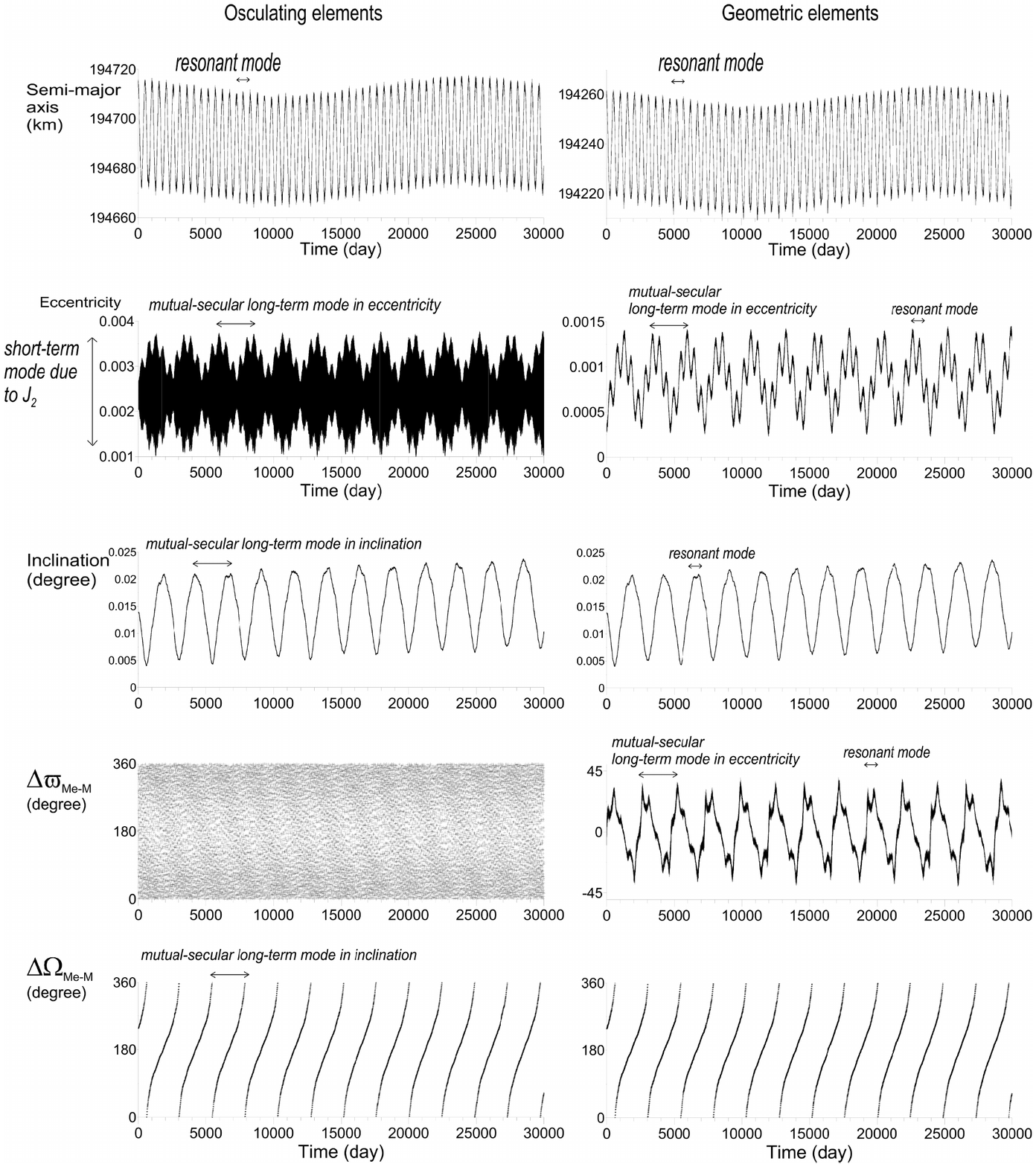}
  \caption{Left and right columns show, respectively, the osculating and corresponding geometric elements (indicated) of a satellite similar to Methone, calculated with numerical schemes i) and II), respectively, described in Section 2; the initial conditions have been taken from \emph{Horizons} system at the date January 01, 2016; see Appendix 1. Note the distinct scales of the y-axes in the plots of the osculating and geometric eccentricity and semi-major axis. The horizontal large arrows, vertical arrows and horizontal small arrows are explained in Section 3.1, 3.2 and 3.3.2, respectively.}
   \label{<meth>}
\end{figure*}

The main goal in this section is to identify the relevant short-term, resonant and long-term components in the time variations of the elements of Methone due to several perturbations. For this task, we utilize numerical simulations which provide us outputs similar to the current orbital elements of Methone\footnote{Recall that ephemerides are limited to a hundred of years, so that numerical simulations are able to cover the long-term components.}. Left and right columns in Fig. \ref{<meth>} show, respectively, the time variations of osculating and corresponding geometric elements: semi-major axis (top), eccentricity (second line), inclination w.r.t. Saturn's equator (third line). Some angular quantities associated are also shown: relative longitudes of pericenters, $\Delta\varpi_{Me-M}\equiv\varpi_{Me}-\varpi_{M}$ (fourth line); and relative longitudes of ascending nodes: $\Delta\Omega_{Me-M}\equiv\Omega_{Me}-\Omega_{M}$ (bottom). All these variables are useful to interpret the orbit of Methone, as shown in the following.

\begin{table*}
 \centering
% \begin{minipage}{140mm}
  \caption{Period and corresponding order of magnitudes of the amplitudes associated to many fundamental frequencies of \emph{osculating} semi-major axis, eccentricity, inclination and respective harmonics. The values have taken from the spectra of a satellite similar to Methone given in Fig. \ref{<meth>}. $P_{\psi_1}$: the 15:14 Mimas-Methone resonance. $P_{\Delta\varpi}$ and $P_{\Delta\Omega}$: mutual secular modes of $\Delta\varpi$ and $\Delta\Omega$, respectively. $P_{4:2}$: the Mimas-Tethys resonance. $P_{VLT}$: very long-term component. $H_1$: third harmonic of $P_{4:2}$. $H_2$ and $H_3$: linear combinations of $P_{\psi_1}$ and $P_{\Delta\varpi}$. The harmonics $H_2$ and $H_3$ will be explained in details in Section 5.1 (see Fig. \ref{<ipsm>}).}
   \vspace{0.5cm}
\begin{tabular}{cccccccc}

\hline
             &  Semi-major axis             &  Eccentricity                & Inclination                     \\
\hline
             &                              &                              &                      \\
Period (day) & $P_{\psi_1}=521.57$          & $P_{\psi_1}=521.57$          &  $P_{\psi_1}=521.57$       \\
Amplitude    & $20$ km                  & $3.7\times10^{-5}$               & $1.2\times10^{-5}$  degree         \\
\\
Period (day) &$P_{4:2}=33,554.43$       & $P_{\Delta\varpi}=2408.21$   &$P_{\Delta\Omega}=2443.28 $    \\
Amplitude    &$4.2$ km                  & $6\times10^{-5}$         & $7.3\times10^{-3}$  degree   \\
\\
Period (day) &$P_{H_1}=11,184.81$           & $P_{H_2}=665.76$             &  $P_{H_4}=1221.64$         \\
Amplitude    & $0.6$ km                 & $9\times10^{-6}$         &  $1\times10^{-3}$ degree    \\
\\

Period (day) &                              & $P_{H_3}\sim428.72$          &  $P_{VLT}=83,886.08$    \\
Amplitude    &                              & $6\times10^{-6}$         &   $2\times10^{-4}$ degree       \\
             &                              &                              &                      \\

\hline
\end{tabular}

%\end{minipage}
\end{table*}

\subsection{Mutual-Secular Long-Term Modes}

We define the ``mutual-secular long-term modes'' as those dictated by $J_2$ secular perturbations, and the Mimas-Methone non-resonant interactions which include secular and long-term perturbations. $J_2$ perturbations contribute significantly to the secular variations of the argument of the pericenter ($\omega$) and the longitude of ascending node ($\Omega$):
\begin{eqnarray}
\dot{{\Omega}}_{J_2}&=&-\frac{3}{2}\frac{J_{2}nR^2}{a^2(1-e^2)^2}\cos i,\label{52}\\
\dot{{\omega}}_{J_2}&=&+\frac{3}{2}\frac{J_{2}nR^2}{a^2(1-e^2)^2}\left(2-\frac{5}{2}\sin^2i\right).\label{53}
\end{eqnarray}
Equations (\ref{52}) and (\ref{53}) are average rates, where $n=\frac{2\pi}{P}$, $P$, $a$, $e$, $R$, $i$ are the mean motion, orbital period, semi-major axis, orbital eccentricity, equatorial radius of the planet and orbital inclination w.r.t. the celestial equator of the planet, respectively.

On one hand, there is no secular variations in eccentricity and inclination due to $J_2$ secular perturbations (e.g. Danby 1988, page 346). On the other hand, the \emph{long-term} Mimas-Methone \emph{non-resonant} mutual perturbations are primary linked with the eccentricity and inclination, which are respectively related to the variations of $\Delta\varpi_{Me-M}$ and $\Delta\Omega_{Me-M}$; and both, in its turn depend on the secular rates of $\omega$ and $\Omega$ due to $J_2$. These \emph{long-term} components appear as regular oscillations of eccentricity and inclination. In Fig. \ref{<meth>}, we indicate by horizontal large arrows the mutual-secular long-term modes in eccentricity and inclinations. In the case of Methone, the accurate values of the corresponding periods associated to these modes are $P_{\Delta\varpi}\sim2408.21$ day and $P_{\Delta\Omega}\sim2443.28$ day, respectively. The spectra given in Fig. \ref{<fft>} show that the highest peaks in eccentricity and inclination are those associated to $P_{\Delta\varpi}$ and $P_{\Delta\Omega}$, respectively. The periods and corresponding amplitudes of the these perturbations are also summarized in Table 1.

\subsection{$J_2$ short-term variations}

$\Delta\Omega_{Me-M}$ circulates in prograde sense with period $P_{\Delta\Omega}$ (Fig. \ref{<meth>}). Note that both, geometric and osculating $\Delta\Omega_{Me-M}$, show this behaviour, since $\Omega_{Me}$ is not affected by $J_2$ short-term variations. For $\Delta\varpi_{Me-M}$ the situation is different: the geometric $\Delta\varpi_{Me-M}$ \emph{oscillates around zero} with relatively large amplitude, while the \emph{osculating} $\Delta\varpi_{Me-M}$ appears like a strip. In the latter case, the long-term oscillatory mode cannot be seen in the plot because it is obfuscated by the $J_2$ short-term component of $\omega_{Me}$ shown in Fig. \ref{<geo>}. The oscillation of the geometric $\Delta\varpi_{Me-M}$ indicates alignment of the orbits of Mimas and Methone at long-term time scales, consisting a fundamental behaviour of the orbit of Methone which will be fully explored in Section 4 (see Fig. \ref{<secular>}).
%(\footnote{We will go back on secular/long-term modes in next section. See Equations (\ref{52}), %(\ref{53}).}.

By comparing the two plots of the eccentricity in Fig. \ref{<meth>}, we can note that the osculating eccentricity suffers short-term variations not present in the plots of geometric eccentricity. Indicated by vertical arrows in Fig. \ref{<meth>}, they are linked to the $J_2$ induced mode in $\omega_{Me}$ discussed in Section 2. The amplitudes of the short-term variations are of the order $\sim10^{-3}$, a value one order of magnitude larger than the associated to the long-term component $P_{\Delta\varpi}$ ($\sim6\times10^{-5}$; Table 1). Short-term $J_2$ perturbations in osculating inclination can be neglected when compared to the case of eccentricity (Fig. \ref{<meth>}).

In the case of the osculating semi-major axis, the main effect of the $J_2$ perturbations is to generate an orbit with (instantaneous) larger semi-major axis than the geometric semi-major axis and larger (instantaneous) osculating eccentricity. The initial values of these quantities at the date 2016-01-01 are: $a_g\approx194,261.57$ km, $a_o\approx194,715.67$ km ($\Delta a\approx454.1$ km); $e_g\approx0.00037$, $e_o\approx0.00267$ ($\Delta e\approx0.0023$), where the subscripts $o$ and $g$ refers to osculating and geometric, respectively, and $\Delta$ indicates their numerical differences.

\begin{table*}
 \centering
% \begin{minipage}{140mm}
  \caption{The arguments of the expanded disturbing function, up to four degree in eccentricities and inclinations, associated to a 15:14 Mean-Motion Resonance (Methone-Mimas). See Table B.4 in Murray and Dermot (1999). In this work, we adopt the equator of Saturn as the reference plane.}
   \vspace{0.5cm}

\begin{tabular}{ccc}
\hline
Critical argument: & Argument of the cosine   \\
(index)            &                          \\
\hline \hline

$\psi_1$ &$15\lambda_{Me}-14\lambda_M-\varpi_M$  \\ \hline

$\psi_2$ &$15\lambda_{Me}-14\lambda_M-\varpi_{Me}$    \\ \hline

$\psi_3$ &$15\lambda_{Me}-14\lambda_M+\varpi_{Me}-2\varpi_M$  \\ \hline

$\psi_4$ &$15\lambda_{Me}-14\lambda_M-2\varpi_{Me}+\varpi_M$ \\ \hline

$\psi_5$ &$15\lambda_{Me}-14\lambda_M+\varpi_M-2\Omega_M$   \\ \hline

$\psi_6$ &$15\lambda_{Me}-14\lambda_M+\varpi_{Me}-2\Omega_M$  \\ \hline

$\psi_7$ &$15\lambda_{Me}-14\lambda_M-\varpi_M-\Omega_{Me}+\Omega_M$   \\ \hline

$\psi_8$ &$15\lambda_{Me}-14\lambda_M-\varpi_M+\Omega_{Me}-\Omega_M$  \\ \hline

$\psi_9$ &$15\lambda_{Me}-14\lambda_M+\varpi_M-\Omega_{Me}-\Omega_M$    \\ \hline

$\psi_{10}$ &$15\lambda_{Me}-14\lambda_M-\varpi_{Me}-\Omega_{Me}+\Omega_M$  \\ \hline

$\psi_{11}$ &$15\lambda_{Me}-14\lambda_M-\varpi_{Me}+\Omega_{Me}-\Omega_M$   \\ \hline

$\psi_{12}$ &$15\lambda_{Me}-14\lambda_M+\varpi_{Me}-\Omega_{Me}-\Omega_M$   \\ \hline

$\psi_{13}$ &$15\lambda_{Me}-14\lambda_M+\varpi_M-2\Omega_{Me}$    \\ \hline

$\psi_{14}$ &$15\lambda_{Me}-14\lambda_M+\varpi_{Me}-2\Omega_{Me}$    \\ \hline

\end{tabular}
%\end{minipage}
\end{table*}

\subsection{Resonant modes}

\subsubsection{\textbf{4:2 Mimas-Tethys quasi-resonant mode}}
Inspection of the plots given at the bottom in Fig. \ref{<meth>} shows us the presence of a complete cycle with very-long variation of the semi-major axis of Methone. It contributes with $\sim4.2$ km in the variation of the Methone's semi-major axis and appears with period of $\sim91.86$ years in simulations (see bottom-left in Fig. \ref{<fft>} and Table 1). This component is related to the proximity of the Mimas-Methone system to the 4:2 Mimas-Tethys inclination type mean-motion resonance, which acts with large amplitude in Mimas elements with a period of $\sim75$ year (see Callegari and Yokoyama 2010a and references). Methone's eccentricity and inclination are not significantly affected by this kind perturbation.

\subsubsection{Corotation mode}
%; the dominant amplitude in this variable are related to the $P_{\Delta\varpi}$ and short-term %variations rather than the resonance. Inclination is weakly affected by nor the Corotation or 4:2 %Mimas-Tethys.

Let us consider now the perturbations due to 15:14 Mimas-Methone Corotation resonance. The semi-major axis of Methone librates with amplitude $\sim20$ km and period $P_{\psi_1}\sim521.57$ day, indicated by horizontal small arrows in Fig. \ref{<meth>}, top. This component shows the signs of the resonance on the orbit of Methone, which are expected to be almost isolated in time variations of the semi-major axis. In fact, semi-major axis has a defined peak in the range of the spectrum related to resonance (indicated by $P_{\psi_1}$ in Fig. \ref{<fft>}). This frequency is also clearly seen in eccentricity plots (small horizontal arrows in Fig. \ref{<meth>}), though the dominant mode is $P_{\Delta\varpi}$ (see Fig. \ref{<fft>}). In the case of the inclination, the amplitude associated to the resonant frequency is very small (Figs. \ref{<fft>}, \ref{<meth>}, Table 1), since in this case we have an eccentricity-type mean-motion resonance.

\begin{figure*}[h]
 \includegraphics[width=12cm]{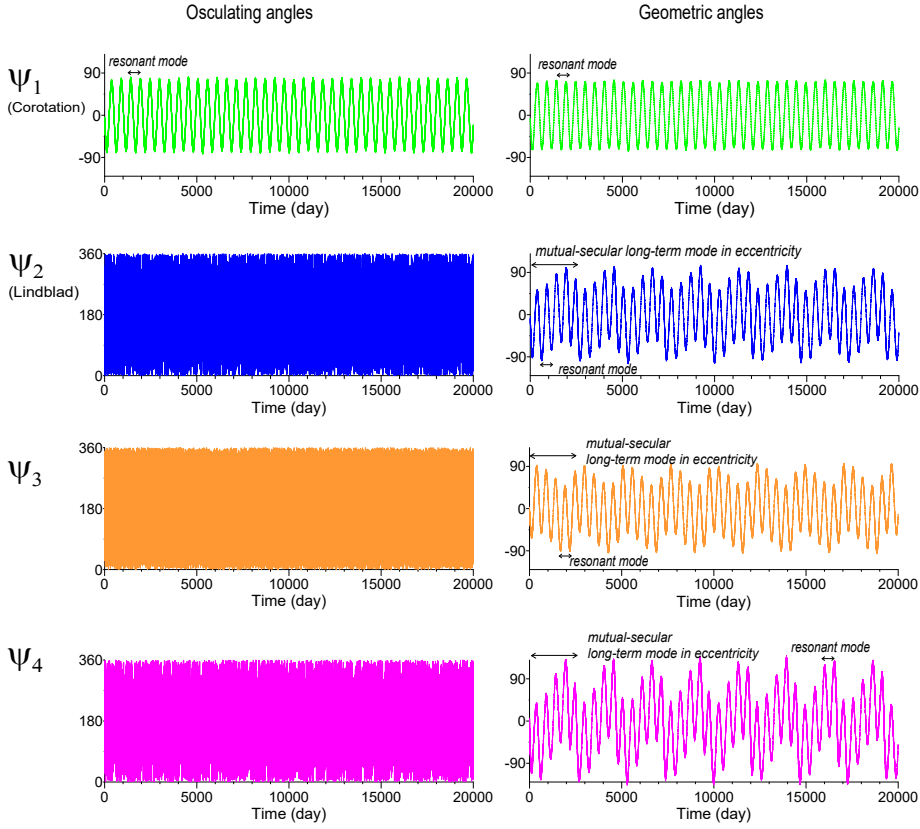}
  \caption{The same as Fig. \ref{<meth>}, where four critical angles are plotted utilizing different colours. $\psi_1=\alpha-\varpi_{M}$, $\psi_2=\alpha-\varpi_{Me}$, $\psi_3=\alpha+\varpi_{Me}-2\varpi_{M}$, $\psi_4=\alpha+\varpi_{M}-2\varpi_{Me}$, where $\alpha\equiv15\lambda_{Me}-14\lambda_M$. In order to highlight the resonant and long-term periods, only the first 20,000 days are shown.}
   \label{<meth-ang>}
\end{figure*}

The real evidence of a resonant system can only be confirmed by the true libration of determined critical angle. Green curves in Fig. \ref{<meth-ang>} show that the corotation angle $\psi_1=15\lambda_{Me}-14\lambda_M-\varpi_{M}$ oscillates around zero with large amplitude of $\sim 68$ degree, obtained from spectrum of $\psi_1$ (not shown for brevity). Thus, Methone is currently trapped into the 15:14 Corotation mean-motion resonance with Mimas. Note that both, geometric and osculating $\psi_1$ librate with very similar amplitudes and periods. This agreement occurs because the corotation angle involves the longitude of pericenter of Mimas ($\varpi_{M}$), as well the mean longitudes $\lambda_{Me}$ and $\lambda_{M}$, three quantities which are not affected by the induced short-term variations due to the $J_2$ perturbations\footnote{This result has been shown in Callegari and Yokoyama (2020) in the case of the satellite Anthe, and the same should be directly applied to any other close-in small satellite of Saturn in Corotation resonance with Mimas.}.

\section{Simultaneous oscillations of four critical arguments}

\subsection{Physical description}
The results given above (Section 3.3.2) regarding the current resonant state of the pair Mimas-Methone are in  agreement with several analyses of the resonance (e.g. El Moutamid et al. 2014). However, motivated by the discussion given in the beginning of Section 1 on the multiple oscillations of resonant arguments, we investigate the resonant perturbations in more details.

By initially adopting a classical procedure devoted to examine resonances implemented in ephemerides-oriented works (e.g. Spitale et al. 2006), we analyze the time variations of angles $\psi_k$, $k=1, ..., 14$ associated to the 15:14 Mimas-Methone mean-motion resonance, as given in Table 2. The results are divided in Fig. \ref{<meth-ang>}, where we show $\psi_1$, $\psi_2$, $\psi_3$, $\psi_4$, and Fig. \ref{<meth2>} in Appendix 2, where we analyze all the others angles.

Curves given at right in Fig. \ref{<meth-ang>} show simultaneous stable long-term oscillations around zero of \emph{four} \emph{geometric} angles $\psi_1$, $\psi_2$, $\psi_3$, $\psi_4$. In order to explain the reason of this phenomena, let us consider initially only the angles $\psi_1$ (corotation) and $\psi_2$ (Lindblad), which can be written through the compact expressions:
\begin{eqnarray}
\alpha&\equiv&15\lambda_{Me}-14\lambda_M, \label{1}\\
\psi_1&=&\alpha-\varpi_{M},\label{2}\\
\psi_2&=&\alpha-\varpi_{Me}\label{3}.
\end{eqnarray}
Inspection of time variation of the geometric Lindblad angle $\psi_2$ (blue curve at the right in Fig. \ref{<meth-ang>}), shows that it is composed clearly by \emph{two main components}: a rapid one, and a long-term mode. The former is associated to the corotation angle $\psi_1$, and the latter to $\Delta\varpi_{Me-M}$. This can be understood after rewriting the expression of $\psi_2$ by adding and subtracting $\varpi_{M}$ in (\ref{3}) as follows:
\begin{equation}
\psi_2=\alpha-\varpi_{Me}=\alpha-\varpi_{M}+\varpi_{M}-\varpi_{Me}=\psi_1-\Delta\varpi_{Me-M},\label{2889}
\end{equation}
where we have utilized Equation (\ref{2}) and $\Delta\varpi_{Me-M}\equiv\varpi_{Me}-\varpi_{M}$. Therefore, since both $\psi_1$ and the geometric $\Delta\varpi_{Me-M}$ oscillate around zero, the result is the plot of the geometric Lindblad angles displayed in Fig. \ref{<meth-ang>}. The time variation of the \emph{osculating} $\psi_2$ circulates due to the rapid component in $\Delta\varpi_{Me-M}$ (Section 3.1).

As stated before, the true angle of the 15:14 Mimas-Methone resonance is the corotation angle. Physically, the libration of $\psi_1$ means that the conjunctions Mimas-Methone repeat in a line oscillating around the pericenter of Mimas. On the other hand, the libration of $\psi_2$ \emph{also around zero} means that the line of the conjunctions of the pair Mimas-Methone oscillates around a reference direction passing through the longitude of pericenter of Methone. Simultaneous oscillation of the geometric $\psi_1$ and $\psi_2$ around zero can only occur in the case of alignment of the geometric orbits of Methone and Mimas. On the contrary, the conjunctions would never periodically repeat close to the directions of both pericenters. Therefore, the interpretation of the simultaneous oscillation of $\psi_1$ and $\psi_2$ (geometric) is the following: the conjunctions Mimas-Methone continue to repeat in a line oscillating in the vicinity the pericenter of Mimas, but due to the second mode in Equation (\ref{2889}), the complete cycle of the conjunctions around a line oscillating around the (geometric) pericenter of Methone closes after a period $P_{\Delta\varpi}\sim2408.21$ day.

%In Figs. \ref{<1514>}(a,b) we draw these two resonant configuration with an artistic %representation\footnote{Fig. \ref{<1514>} shows a fictitious scenario, i.e., where the semi-major axes are %not given in scale and the orbital eccentricities are exaggerated. In the real case, the proper %eccentricity and inclination of Methone (perturbed body), are always smaller than those of the Mimas (main %disturber). The outer satellite must have negligible mass with respect to the inner's one in order to this %scenario be applied to a pair like Mimas-Methone.}.

%Fig. \ref{<1514>}(c) illustrates that the

%\footnote{Note also that $\psi_1$ does not contain the long-term variation, at lest the accuracy of our %analyses.}

%Note that, since both $\psi_1$ and $\psi_2$ librate, we would be tented to choice as a fundamental frequency of the system the angle $\psi_2$, so that $\psi_1=\psi_2+\Delta\varpi_{Me-M}$, from (\ref{2889}). Having in mind that simultaneous oscillation occurs due to aligned orbits (i.e., oscillation of the geometric $\Delta\varpi_{Me-M}$ around zero), in the practice the result would be the same.

%Thus, in this case the libration $\psi_2$ would contain two modes, and with this choice $\psi_2$ would %not be a \emph{pure mode} representing the fundamental frequency of the resonant system.

\vspace{1cm}
Let us consider now the cases of $\psi_3$ and $\psi_4$. From Table 2, we have:
\begin{eqnarray}
\psi_3&=&\alpha+\varpi_{Me}-2\varpi_{M},\label{388}\\
\psi_4&=&\alpha+\varpi_{M}-2\varpi_{Me}.\label{488}
\end{eqnarray}

From (\ref{388}), we obtain:
\begin{eqnarray}
\psi_3&=&\alpha+\varpi_{Me}-\varpi_{M}-\varpi_{M},\nonumber\\
      &=&\underbrace{\alpha-\varpi_{M}}+\underbrace{\varpi_{Me}-\varpi_{M}},\nonumber\\
      &=&\hspace{0.5cm}\psi_1 \hspace{0.4cm}        +\Delta\varpi_{Me-M}.\label{5}
\end{eqnarray}

Analogously, in the case of (\ref{488}), we have:
\begin{eqnarray}
\psi_4&=&\alpha+\varpi_{M}-\varpi_{Me}-\varpi_{Me},\nonumber\\
      &=&\underbrace{\alpha-\varpi_{M}}+\varpi_{M}-\varpi_{Me}-\varpi_{Me}+\varpi_{M},\nonumber\\
      &=&\hspace{0.5cm}\psi_1 \hspace{0.3cm}        -(\varpi_{Me}-\varpi_{M}+\varpi_{Me}-\varpi_{M})\nonumber\\
      &=&\psi_1         -2\Delta\varpi_{Me-M}.\label{6}
\end{eqnarray}
Equations (\ref{5}) and (\ref{6}) show that $\psi_3$, $\psi_4$ can also be written as combinations of $\psi_1$ and $\Delta\varpi_{Me-M}$. So, the geometric $\psi_3$, $\psi_4$ also oscillates around zero.

In Appendix 2, additional similar analyses are given in the cases of all other geometric angles $\psi_k$, $k\geq5$, where all of them circulate.

We conclude that, since $\psi_1$ librates around zero and the geometric $\Delta\varpi_{Me-M}$ also oscillates around the same center, the simultaneous oscillations of $\psi_1$, $\psi_2$, $\psi_3$, $\psi_4$ is thus explained.

%\begin{figure}
%  \resizebox{\hsize}{!}{\includegraphics{1514.jpg}}
%     \caption{\textbf{(a) }Fictitious orbital configuration of two satellites trapped into Corotation %mean-motion resonance. Symbols 1 and 2 indicate the positions of repeated conjunctions belonging to a line %(dashed line) which oscillates around a direction passing though of the longitude of  pericenter of the %inner satellite (the latter is indicated by full square symbol). \textbf{(b)} The same as (a) in the case %of a kind of Lindblad resonance, where conjunctions occur in a line oscillating around the longitude of %pericenter of the outer satellite (represented by triangle). \textbf{(c)} The fictitious situation of %simultaneous oscillation into Corotation and Lindblad mean-motion resonances. See also footnote number %6.}
%     \label{<1514>}
%\end{figure}

 \begin{figure*}
\centering
   \includegraphics[width=12cm]{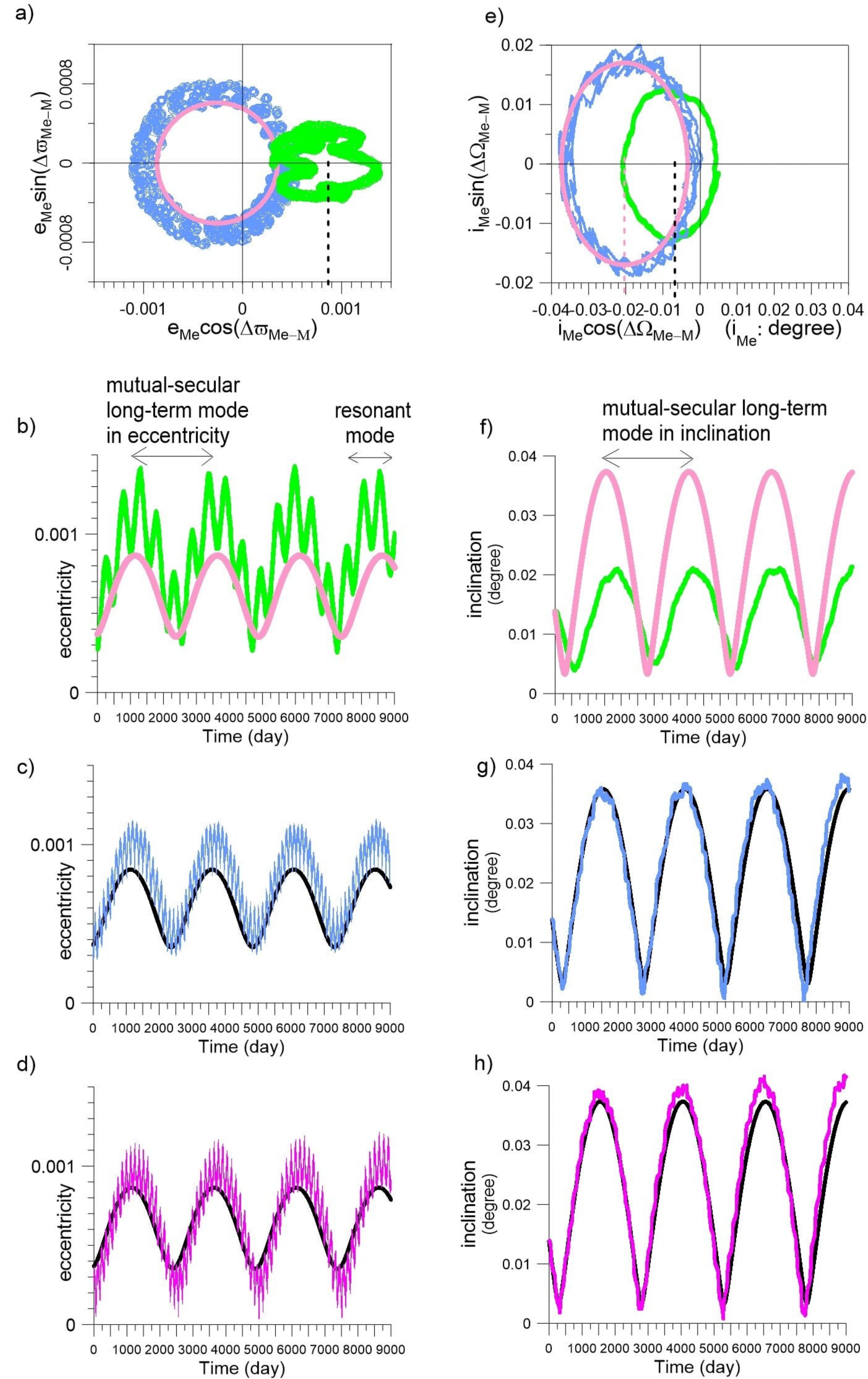}
     \caption{Green lines: \emph{geometric elements} taken from numerical simulation given in Fig. \ref{<meth>}. Pink and black curves: obtained from classical secular theory including an oblate Saturn with $J_2$ and $J_4$, and Mimas (see Section 4.2). Light-blue and magenta lines: the same as the green lines, except that the initial osculating semi-major axes are $a_0=194,775$ km (blue) and $a_0=194,620$ km (magenta).
     \textbf{(a)} Projection onto the plane $(e_{Me}\cos(\Delta\varpi_{Me-M}),e_{Me}\sin(\Delta\varpi_{Me-M}))$. The vertical black dashed line indicates the forced values of the geometric eccentricity.
     \textbf{b)} Time variation of the eccentricity corresponding to (a). The main long-term and resonant modes are indicated by arrows.
     \textbf{(c,d)} Time variations of the eccentricities.
     \textbf{(e-h)} Plots of the orbital inclinations (w.r.t. Saturn equator), corresponding to all cases given at left column. In (e) the orbit is projected onto the plane $(i_{Me}\cos(\Delta\Omega_{Me-M}),i_{Me}\sin(\Delta\Omega_{Me-M}))$, and the vertical pink dashed line indicates the forced inclination of a non-resonant test satellite.}
     \label{<secular>}
\end{figure*}

\subsection{The forced eccentricity and inclination of Methone}

\subsubsection{The forced eccentricity of Methone}

In this section, we will provide evidence that the alignment of the geometric orbits of Mimas and Methone, represented by the oscillation of $\Delta\varpi_{Me-M}$ around zero, is related to the forced eccentricity of Methone due to the 15:14 Mimas-Methone resonance.

Fig. \ref{<secular>}(a) displays the projections of the elements of Methone on the plane ($e_{Me}\cos(\Delta\varpi_{Me-M}),e_{Me}\sin(\Delta\varpi_{Me-M})$). Green curve shows a forced component in the geometric eccentricity of $\sim0.0008$, while the proper mode is almost a half of this value\footnote{Recall that the forced eccentricity, roughly indicated by vertical dashed line, is measured from the centre of the coordinate system, while the proper mode is taken from the forced centre.}. This is the reason why the geometric $\Delta\varpi_{Me-M}$ oscillates around zero.

Green curve in Fig. \ref{<secular>}(b) shows the time variation of the eccentricity given in Fig. \ref{<secular>}(a). This plot highlights the two main frequencies which compose the numerical solutions: the higher frequency associated to the 15:14 Mimas-Methone Corotation resonance \textbf{(period $\sim 521.57$ day)}, and the slower one associated to mutual-secular long-term mode (\textbf{period $\sim2408.21$ day}). Note that the green curve is a detailed view of the first 9000 days of the eccentricity plot shown in Fig. \ref{<meth>} and interpreted in Sections 3.1 and 3.3.2.

We can ``isolate'' the long-term components of the geometric eccentricity by solving the linear solutions of the generalized Laplace-Lagrange secular theory. For this task, we utilize linear equations given in section 7.7 in Murray and Dermott (1999) for a system formed by an oblate Saturn with $J_2$ and $J_4$, Mimas, and a particle with the same initial conditions of the geometric orbit of Methone. These equations are valid for second order in eccentricity and inclination, which is a good approximation in the case of Mimas-Methone system. The result is the pink curve in Fig. \ref{<secular>}(b). We can note that the period obtained with linear theory shows good agreement with the current values (that with period $P_{\Delta\varpi}$). Pink curve in Fig. \ref{<secular>}(b) has also been projected on the plane in Fig. \ref{<secular>}(a). By comparing the pink and green curves in Fig. \ref{<secular>}(a), we have that the forced geometric eccentricity obtained from the linear theory does not match the current values ($\sim0.0008$).

From the above calculations we can infer that forced geometric eccentricity of Methone has an additional component not contemplated by secular linear theory. In the following we will give numerical evidences that the forced component is related to the 15:14 Mimas-Methone mean-motion Corotation resonance. Consider an orbit very close to the current Methone, but not trapped into resonance. In this case, the forced eccentricity of the projected elements would be well-determined by linear secular theory. Blue curve in Fig. \ref{<secular>}(a) corresponds to an orbit with initial semi-major axis $a_0=194,775$ km, differing $\sim59.3$ km of the current value $a_{Me}=194,715.67$ km\footnote{The other initial elements in the simulation are the same than those of Methone at the date 2016-01-01; $a_0=194,775$ km has been taken from Fig. \ref{<ipsm>}. As discussed in item i) in Section 5.1, at this initial conditions, the test particle is sufficiently far enough of the chaotic layers of the Corotation zone.}. As conjectured above, the forced eccentricity in the current orbit of Methone ($\sim0.0008$) is no more present.

Fig. \ref{<secular>}(c) shows the time variations of the geometric eccentricity (blue curve), and the eccentricity calculated with linear theory for this orbit (black line). Blue curve is composed basically by rapid variations driven by the long-term component. The rapid oscillations are associated to the proximity of the system to the 15:14 resonance (a state which can be considered a ``quasi-resonance'').

In Fig. \ref{<secular>}(d) we give another example $a_0=194,620$ km. This initial condition is located at opposite ``side'' of the Corotation resonance, as we will see in Section 5.1. The results are very similar to Fig. \ref{<secular>}(c).

\subsubsection{The forced inclination of Methone}

Figures at right column in Fig. \ref{<secular>} show plots of the orbital inclinations corresponding to all simulations given at left column.

Green curve in Fig. \ref{<secular>}(e) displays the projections of the elements of Methone on the plane ($i_{Me}\cos(\Delta\Omega_{Me-M}),i_{Me}\sin(\Delta\Omega_{Me-M})$). The forced inclination of Methone is $\sim0.007$ degree, and the forced centre is indicated by vertical black dashed line. Since the proper geometric inclination of Methone is larger than the forced one, the green curve encompasses the origin. Therefore, $\Delta\Omega_{Me-M}$ circulates for the current orbit of Methone with period $P_{\Delta\Omega}$ (with starting time at the date 2016-01-01). Note also that the amplitude of the proper mode in inclination is dominated by long-term mode since perturbations of the Corotation resonance in inclination has negligible amplitude, as we already have noted in Fig. \ref{<meth>}, Sections 3.1 and 3.3.2. See also Fig. \ref{<secular>}(f).

The pink curve in Fig. \ref{<secular>}(e) shows the mutual-secular long-term mode in inclination obtained with linear secular theory. $\Delta\Omega$ taken from this simulation oscillates around $\pi$, and the forced centre is located at $\sim0.02$ degree (and it is indicated by vertical pink dashed line).

At this point we can conclude that the forced term in inclination of Mimas due to Corotation resonance contributes with the current orbit of Methone by precluding the anti-alignment of their ascending nodes (or, in other words, temporary oscillations of $\Delta\Omega_{Me-M}$ around $\pi$\footnote{Evidently, this conclusion is valid within in timescale utilized in this work (limited to tens of thousands of years).}). At first glance, this is a surprising result once we recognize the Corotation resonance as an ``e-type'' resonance. However, inspection of Table B.4 in Murray and Dermott (1999) shows us that one of the terms of the expanded disturbing function associated to Corotation resonance is proportional to the square of inclination of Mimas. An in-deep investigation if this particular term of the perturbing potential can force the plane of orbit of Methone due to the relatively large value of Mimas' inclination ($\sim1.56$ degree) is, however, out from the main goals of this project.

As we have done in the case of the eccentricity of Methone, it is possible to measure the role of the Corotation resonance in the inclination of Methone by analyzing individual test satellites with initial conditions located outside the Corotation resonance. Again we utilize the numerical simulation given in blue curves in Fig. \ref{<secular>}. Due to forced component in inclination, $\Delta\Omega$ may episodically oscillate around $\pi$. When the numerical integration is extended to longer time, the projected trajectory occupy all quadrants in Fig. \ref{<secular>}(e) and $\Delta\Omega$ circulates.

 %contain only the long-term mutual (plus $J_2$) components.
%The plot of the geometric eccentricity of Methone given in Fig. \ref{<secular>}(c) highlights two main frequencies which compose green curve in (a): the higher frequency associated to the resonant 15:14 mean-motion perturbations (period $P_{\psi_1}\sim521.57$ day), and the slower one with $P_{\Delta\varpi}\sim2419.79$ day (i.e., the long-term mode associated to mutual Mimas-Methone secular perturbations and $J_2$).

We have therefore numerically shown that: i) The forced eccentricity due to Corotation resonance explains the long-term alignment of the orbits of Mimas and Methone in the numerical simulations; ii) The amplitudes of the forced modes of eccentricity and inclination of Methone's orbit are primary dictated by the 15:14 Mimas-Methone Corotation mean-motion resonance.

\begin{figure*}
\centering
   \includegraphics[width=6cm]{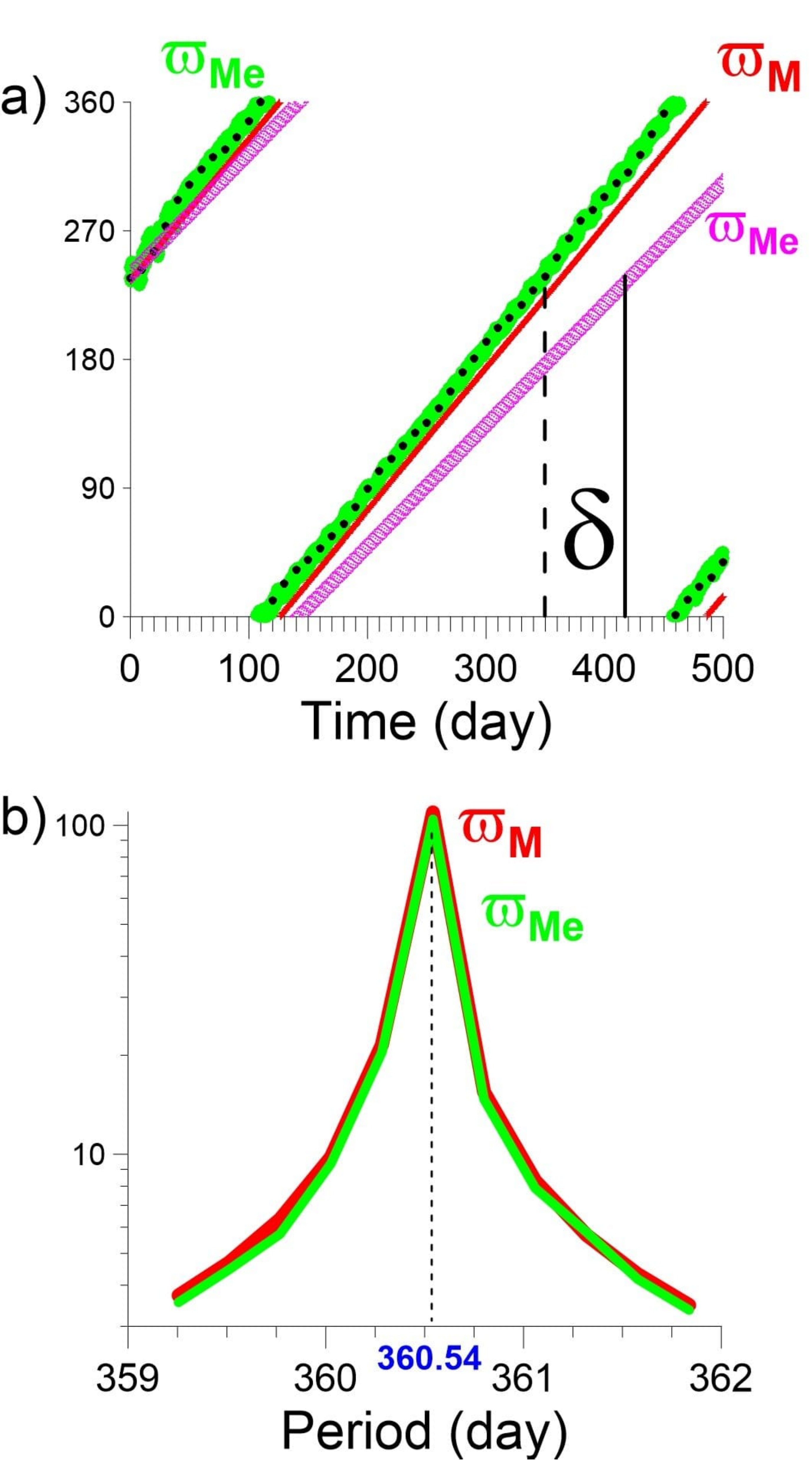}
   \caption{\textbf{(a)} Time variations of the geometric longitudes of pericenters of satellites similar to Methone ($\varpi_{Me}$) and Mimas ($\varpi_M$), obtained with different simulations S1 (green and red lines), S2 (magenta), and S3 (black points). Dashed and full vertical lines indicate the first cycle of circulation of $\varpi_i$; $\delta$ is the difference of $\sim63.5$ day discussed in the text. \textbf{(b)} Corresponding Fourier spectra of the same variables given \textbf{in (a) taken from S1}. Only the details in the short interval around the main peaks centered at $\sim 360.5419$ day (highlighted by dotted line) have been shown.}
   \label{<espectroa>}
\end{figure*}

\subsection{The precession rates of $\varpi_{Me}$ and $\varpi_{M}$}
%The stationary state of the geometric $\Delta\varpi_{Me-M}$ is stationary is that simultaneous variations of the geometric longitude of pericenter of Methone are similar to the precession of the pericenter of Mimas.

The geometric $\Delta\varpi_{Me-M}$ is almost stationary because the precession rates of the geometric $\varpi_{Me}$ and $\varpi_{M}$ have the same order of magnitude. Thus, a careful analysis of these two latter variables should be useful in order to quantify the main long-term component in eccentricity. For this goal, we have performed a large set of individual numerical simulations. The main conclusions are resumed by simulations S1, S2 and S3 shown in Fig. \ref{<espectroa>}. Simulation S1 is obtained with numerical scheme i), where $J_2$, $J_4$ of Saturn and the satellites Mimas, Methone, Enceladus, Tethys, Dione, Rhea and Titan have been included. The initial conditions have been taken from \emph{Horizons} system of ephemerides in the date January 01, 2016. The total time of integration is 1720 year, with output of 0.06 day. S2 is similar to S1, but only Methone and the zonal gravity harmonics have been considered. S3 includes Mimas, Methone and the zonal harmonics.

Vertical dashed line in Fig. \ref{<espectroa>}(a) give the first cycle of circulation of the geometric $\varpi_{Me}$ (green curve) and the geometric $\varpi_M$ (red curve) taken from the numerical simulation S1. From Fig. \ref{<espectroa>}(a), we have that $\varpi_{Me}$ and $\varpi_M$ complete the circulations at approximately the same period.

Fig. \ref{<espectroa>}(b) shows the spectra of $\varpi_{Me}$ and $\varpi_{M}$ taken from S1. The peaks given in the figure are the largest ones present in the spectra of the geometric $\varpi_M$ and $\varpi_{Me}$, defining therefore the periods of their precession rates. Both peaks are centered at $\sim 360.54$ day, showing that they have virtually the same value as well similar amplitudes.
%in the simulation, just as they are the amplitudes (109.5669 and 104.3329, respectively).

Suppressing all satellites and keeping only a body similar to Methone revolving around an oblate Saturn (simulation S2 - magenta line in Fig. \ref{<espectroa>}(a)), $\varpi_{Me}$ circulates with a period $\sim424.0240$ day (obtained from spectrum of S2, not shown), a value slightly larger than the current $360.5419$ day by an amount of $\delta\sim63.5$ day. If we remove all satellites but the pair Mimas-Methone, the geometric $\varpi_{Me}$ (respect to simulation S3, black points in Fig. \ref{<espectroa>}(a)), superposes the green line corresponding to the variation of the geometric $\varpi_{Me}$. The combined conclusions on simulations S2 and S3 are: i) In the particular case of the variation of the pericenter, the disturbances of the other satellites can be neglected when compared to those of Mimas; ii) Mimas is responsible by the acceleration of $\varpi_{Me}$, making its frequency very similar to that $\varpi_M$.

%The exact correspondence of the current long-term precession rates of $\varpi_M$ and $\varpi_{Me}$ is a surprising result since the orbits of Methone and Mimas are separated by $\sim8691$ km, and it is well known that the long-term precession rates depends essentially on the distance in the case of quasi-circular and planar orbits.

%(\ref{59}) is close to the current long-term period circulation of $\varpi_{M}$ ($\sim362.359$ day), and the difference of $7.27$ day is due to the other mid-sized satellites;

%We can show analytically that there would have a difference of this order of magnitude (namely, $\sim67.7$ day) in the precession rates of Mimas and Methone in the case where all perturbations but $J_2$ are neglected. From Equations (\ref{59}), (\ref{60}) we obtain the value $\sim67.7$ day.
%\footnote{(\ref{60}) differs from the current precession rate ($\sim362.359$ day) by $\sim71$ day, a value %almost the same obtained numerically above ($\delta\sim65$ day).}.

%\footnote{Remarks: i) Results from S2 (open circles in Fig. \ref{<espectroa>}(c)), show that the disturbances of all satellites can be neglected when compared to $J_2$ and Mimas; ii) $J_4$ perturbations are equally negligible (Callegari and Yokoyama 2020); iii) iv) Peak 4 in Fig. \ref{<espectroa>}(a) is the component of $\Omega_{Me}$ ($\varpi=\omega+\Omega$). The period associated to the peaks 4 and 9 (Fig. \ref{<espectrob>}(a); Appendix 2) are the same.}.

 \begin{figure*}
 \includegraphics[width=12cm]{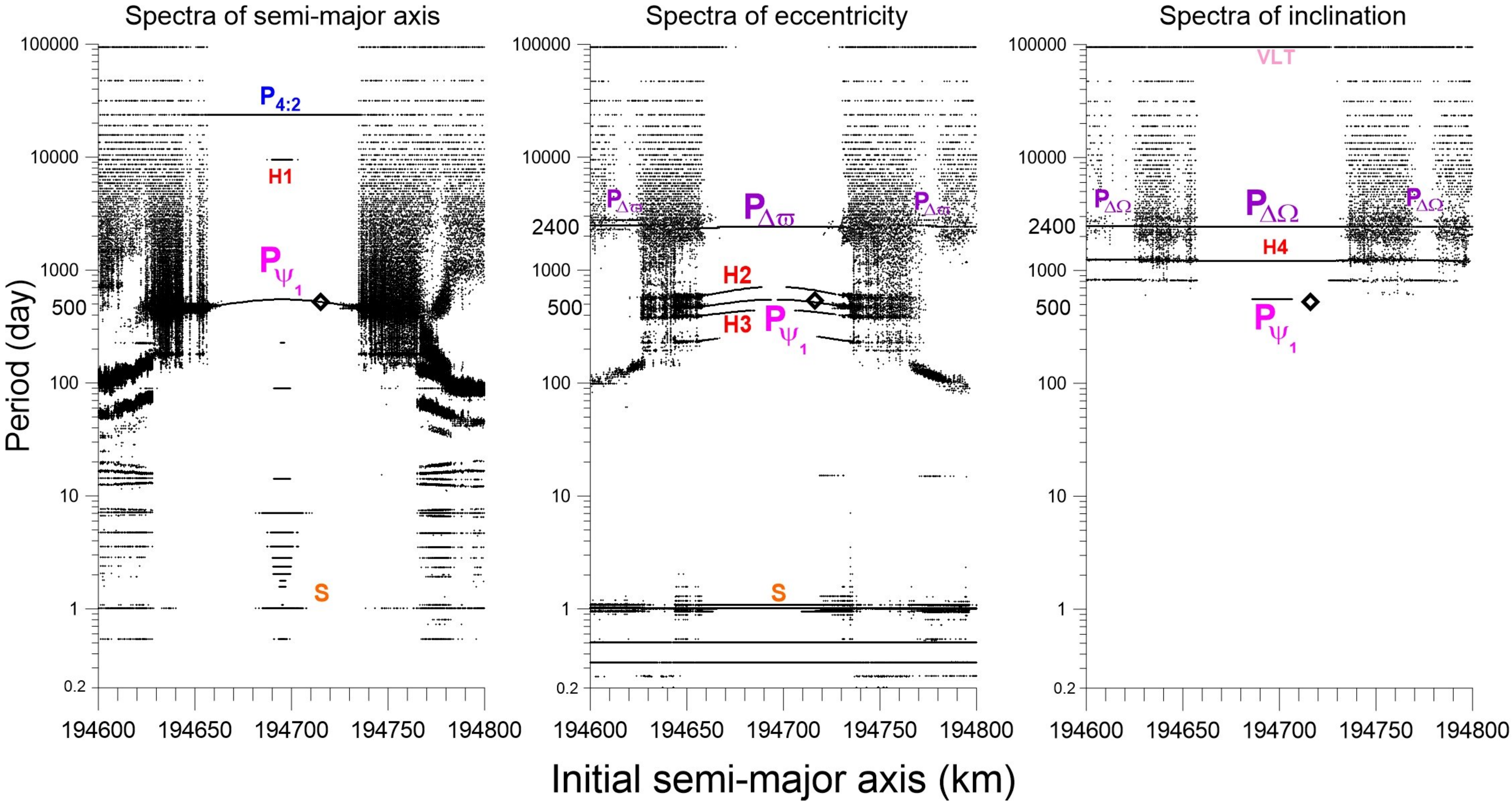}
  \caption{Individual Power Spectrum (IPS) of 1100 clones of Methone constructed from spectra of distinct variables indicated at the top of the plots. The reference amplitude is $5\%$, and y-axes are given in logarithmic scale. The time of integration of each initial state is 258.376 year, and the data have been sampled each 0.09 day. The model for numerical simulations includes the perturbations of $J_2$, $J_4$, Mimas, Enceladus, Tethys, Dione, Rhea and Titan (numerical scheme i)), and the initial conditions have been taken from \emph{Horizons} system of ephemerides at the date January 01, 2016, 00:00. The current configuration of Methone in the plots are indicated by open diamonds. The loci of the fundamental periods are distributed in the vertical direction, and identified with different symbols: \textbf{$P_{\psi_1}$}: the 15:14 Mimas-Methone Corotation resonance. \textbf{$P_{\Delta\varpi}$}, \textbf{$P_{\Delta\Omega}$}: Mutual-secular long-Term Mimas-Methone perturbations. \textbf{$P_{4:2}$}: component of the 4:2 Mimas-Tethys resonance (centered at $\sim33,555.43$ day$=91.87$ year. \textbf{S}: Short-term perturbations. \textbf{VLT}: Very long-term perturbations in inclination and eccentricity. \textbf{H1}, \textbf{H2}, \textbf{H3} and \textbf{H4} and the loci of harmonics between distinct fundamental frequencies. See Section 5.1 for a detailed description of the exact combinations of the harmonics, and Table 1 for accurate amplitudes and periods.}
  \label{<ipsm>}
\end{figure*}

\section{Dynamical mapping of the phase space of the 15:14 Mimas-Methone Mean-Motion Resonance}

In previous section, we provided a detailed description of the dynamics of Methone considering specific orbital configuration of the orbit of the small satellite and the other bodies. Methone is trapped in the 15:14 resonance with Mimas, and our goal now is to explore numerically the domains of the phase space of the mean-motion resonance. We will show resonance mappings constructed in the frequency domain through Fourier spectra of the numerically integrated ensembles of orbits with initial states in the close vicinity the real orbit of Methone.
%The spectra have been obtained with the FFT algorithm given in Press et. al (1996).

The mapping is done in two steps: first, we make an unidirectional sweeping of the resonance by adopting in the numerical simulations the initial semi-major axis of test satellites similar to Methone, $a_0$, as a free parameter. The results will be the so-called ``individual dynamic power spectra'', hereafter denoted by IPS. After that, we add a new dimension, namely the orbital eccentricity $e_0$, resulting in a two-dimensional dynamical map.

While the simulations are performed in dense sets of initial values of $a_{0}$ (IPS), or $(a_0,e_0)$ (dynamical maps), the other orbital elements of the test satellites are set equal to the Methone's at the date January 01, 2016. The larger satellites have their initial states fixed at the same date in all simulations. For each initial condition, the osculating semi-major axis, eccentricity and inclination are analyzed in the Fourier domain (see Section 2). The y-axis of an IPS shows, for each initial condition, the periods associated to the peaks in the spectrum with amplitudes larger than a prefixed fraction. We will denote this threshold by the reference amplitude (hereafter denoted by RA), which usually is somewhat like $1\%$ or $5\%$ of the largest peak. Dynamical maps are based on the spectral number, $N$, defined as the \emph{number} of significant peaks in the spectrum, i.e., those which are above the RA. A colour scale in the map is associated to values of \emph{N} as follows: yellowish/dim colours corresponds to small/large $N$, respectively. For $N>N'$, where $N'$ is a prefixed value of $N$, black colour is applied, so that we have a superior saturation of the scale.

The adopted values of RA in both IPS and dynamical maps given above are in some sense arbitrary, and the choice must always highlight the peaks with physical significance in that spectrum of determined variable. It is important to note that, in the determination of the RA, all peaks linked to the high frequencies are not taken into account, so that amplitudes associated to periods smaller than 15 days are neglected. (15 day is the orbital period of the most distant satellite included in the simulations, namely, Titan.) Therefore, we have a kind of filtering in the frequency domain since we exclude in the determination of the RA all amplitudes accounting for all short-term perturbations in the spectra (including the $J_2$ short term ones).

\subsection{Individual Dynamic Power Spectra}

Fig. \ref{<ipsm>} shows three IPSs of 1100 test satellites calculated from the spectra of semi-major axis (left), eccentricity (middle), and inclination (right). IPSs are very useful to map the resonance domains at a first glance. In fact, the isolated lines indicated by \textbf{$P_{\psi_1}$} in these plots give the dependence of the period of libration of $\psi_1$ with $a_0$, where $\psi_1$ is the corotation angle $\psi_1=15\lambda_{Me}-14\lambda_M-\varpi_{M}$. $\psi_1$ librates around zero in the interval $194,655\leq a_0\leq 194,735$ km. Note that the variation of $P_{\psi_1}$ with $a_0$ is small. Thus, in this interval of semi-major axis, the dominant peak in the spectra is the one associated to the resonance, in a similar way to that shown in Fig. \ref{<fft>}, where $a_0=194,715.67$ km and $P_{\psi_1}\sim521.57$ day (indicated by open diamond in IPSs).

The curve \textbf{$P_{\psi_1}$} is interrupted at the boundaries of the resonance in the IPSs. At these initial conditions, the orbits are non-regular, being certainty chaotic due to the fact that they can belong to the separatrices of the resonance. This explains the aspect of the IPS at these regions, since the spectra are filled with aleatory amplitudes instead to be associated to fundamental ones. These latter conclusions can be assured from several analyses of individual trajectories taken as initial conditions at the borders of the resonance domain (we will return to this point in next Section 5.2, Fig. \ref{<chi>}.)
%(In the vicinity of the orbit of Anthe, $P_{\Delta\varpi}\sim1797.6$ day.)

Many other conclusions can be obtained from IPS regarding on the dynamics of Methone and confronted with the results of Section 3:

i) Symbols \textbf{$P_{\Delta\varpi}$}, \textbf{$P_{\Delta\Omega}$} in IPSs of the orbital eccentricity and inclination show, respectively, the loci of the period of the mutual-secular long-term Mimas-Methone modes in eccentricity and inclination. Their periods are almost constant in the small interval of semi-major axis considered in Fig. \ref{<ipsm>}, what is an expected result due to the nature of these perturbations, as discussed in Section 3.1. \textbf{$P_{\Delta\varpi}$}, \textbf{$P_{\Delta\Omega}$} also appear for semi-major axes far from the Corotation zone\footnote{Exactly over the boundaries of the Corotation resonance, the continuations are broken, which is another possible indicator of chaoticity at the separatrices.};

ii) There are two harmonics, denoted by \textbf{H2} and \textbf{H3}, distributed around the curve \textbf{$P_{\psi_1}$} in IPS of eccentricity. They are linear combinations of $P_{\psi_1}$ and $P_{\Delta\varpi}$, so that \textbf{H2} and \textbf{H3} are the combinations $f_{\psi_1-\Delta\varpi}$ and $f_{\psi_1+\Delta\varpi}$, respectively, whose corresponding periods are $P_{H_2}=P_{(\psi_1-\Delta\varpi)}$, $P_{H_3}=P_{(\psi_1+\Delta\varpi)}$\footnote{For instance, by adopting $P_{\Delta\varpi}=2408.21$ day and $P_{\psi_1}=521.57$ day, we have $P_{(\psi_1+\Delta\varpi)}=428.718$ day from the expression: $\frac{1}{P_{(\psi_1+\Delta\varpi)}}=\frac{1}{P_{\psi_1}}+\frac{1}{P_{\Delta\varpi}}$.}. Another harmonic, denoted by \textbf{H4} in IPS of inclination, has exactly a half of the period of $P_{\Delta\Omega}$. \textbf{H2}, \textbf{H3} and \textbf{H4} have been identified early in spectra of the eccentricity and inclination given in Fig. \ref{<fft>} and Table 1;

iii) Symbol \textbf{$P_{4:2}$}, centered at $\sim33,555.43$ day$ = 91.87$ year is the component of the 4:2 Mimas-Tethys resonance at the semi-major axis of the test satellites, including the current Methone. This perturbation has been previously discussed in Section 3.3.1. IPS of semi-major axis and Fig. \ref{<fft>} also show the presence of a pure harmonic of  $P_{4:2}$ with one third of the fundamental period (denoted by \textbf{H1});

iv) Symbol \textbf{S} shows the short-term peaks in the spectra. The IPS of inclination is free from these perturbations, at least with the adopted RA of $5\%$, reflecting the fact that the inclination of the test satellites are not significantly affected by the short-term variations due to $J_2$, as occurs in the case of eccentricity (as highlighted in the IPS of eccentricity and discussed in Sections 1 and 3.2). These short-term perturbations are enhanced in the spectra of semi-major axis at initial condition close to the center of the resonance, where the amplitude associated to the libration of $\psi_1$ tends to zero;

vi) Symbol \textbf{VLT} are perturbations with very long-term which affect mainly the inclination within the resonance zone. This kind of perturbations, with periods of hundreds of years or more, may be important to the general problem of long-term stability of small satellites, as already discussed in Callegari and Yokoyama (2010b). However, in this work, their complete characterization is beyond our main goals, which are focused on the resonant properties of the Mimas-Methone pair.

 \begin{figure*}
 \includegraphics[width=12cm]{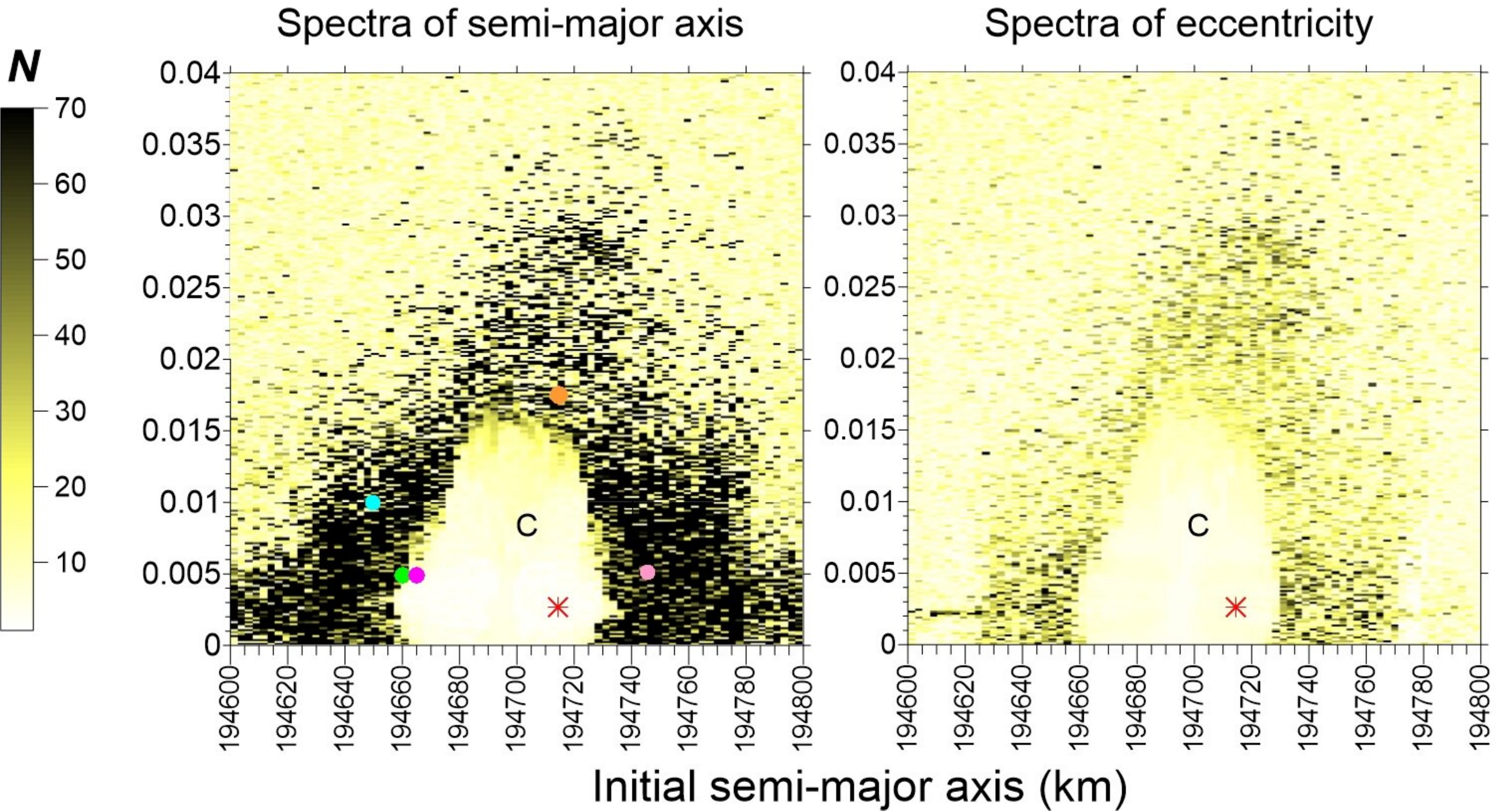}
  \caption{Dynamical maps constructed from the spectra of the osculating semi-major axis (left) and orbital eccentricity (right) of 28,281 clones of Methone. $N$ is the spectral number defined in Section 5.1 with reference amplitude of $5\%$. \textbf{C} is the Corotation zone associated to the 15:14 Mimas-Methone resonance. The system of equations and initial conditions are the same utilized in Fig. 1, except that Rhea and Titan have not been included in the numerical simulations. For each initial condition, the total integration time of numerical simulation is 188,743.68 day ($\sim 516.75$ year), sampled each 0.18 day. The initial elements of Methone are indicated by the red star. Different full discs correspond to initial conditions ($a_0,e_0$) of the orbits shown in Fig. \ref{<chi>}: Orange: (194,715,0.017); Cyan: (194,650,0.01); Green: (194,660,0.005); Magenta: (194,655,0.005); Pink: (194,745,0.005) (semi-major axes in km).}
  \label{<dmm>}
\end{figure*}

 \begin{figure*}
 \centering
 \includegraphics[width=12cm]{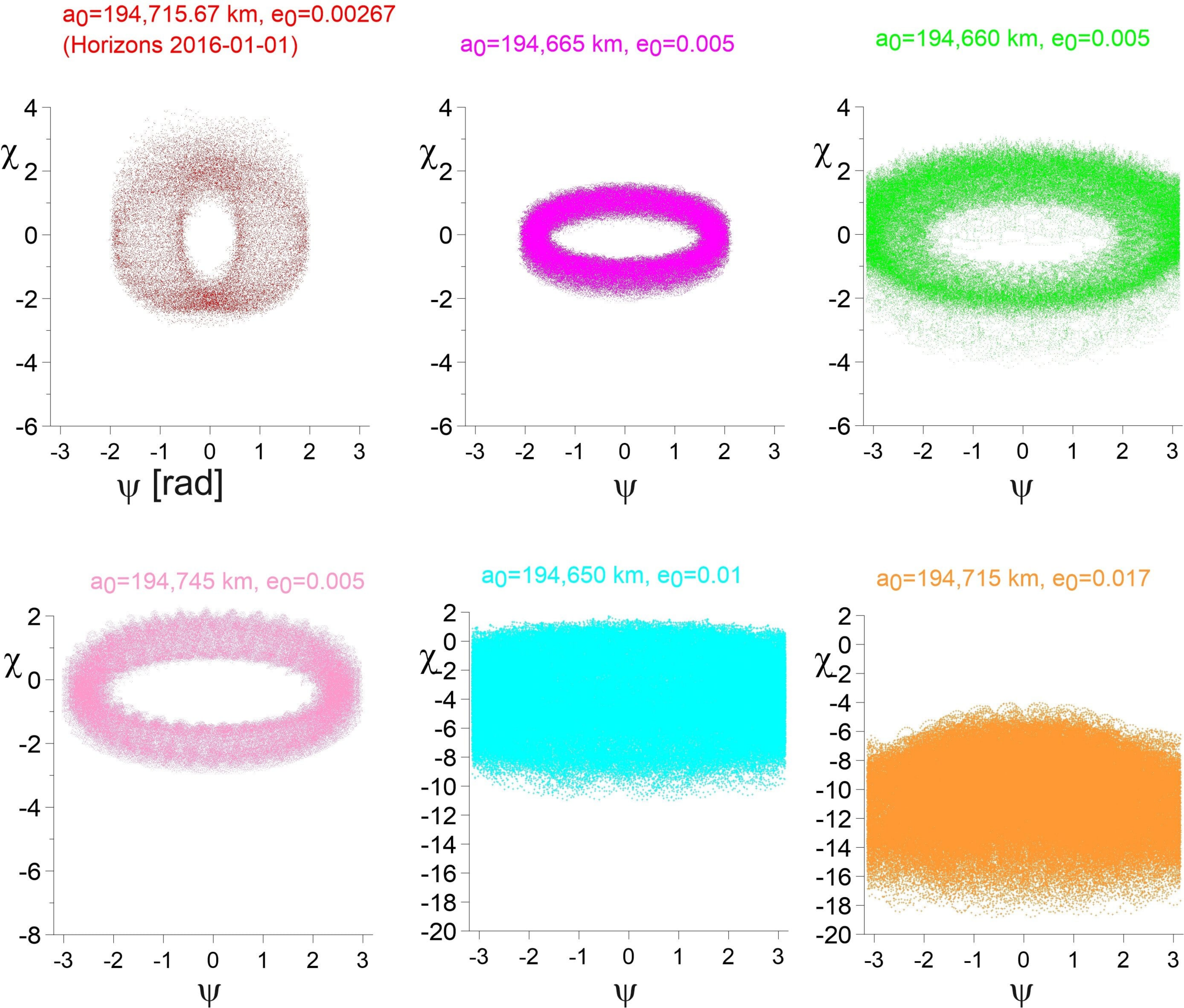}
  \caption{Plots of orbits projected onto the plane $(\chi,\psi)$ with initial conditions given in coloured full discs in Fig. \ref{<dmm>}. Inside the Corotation zone, $\chi$ represents the width of the resonance with the associated angle $\psi=15\lambda_{s}-14\lambda_M-\varpi_{M}$, where $s$ is a test satellite. See the main text and footnote 10 for details.}
  \label{<chi>}
\end{figure*}

\subsection{Dynamical Maps}

The initial osculating eccentricity of the test satellites considered in all simulations in Fig. \ref{<ipsm>} is $e_0=0.00267$, corresponding to the value of Methone at the date January 01, 2016 in \emph{Horizons} system of ephemerides. Let us now to investigate how the width of the resonance depends on $e_0$. Fig. \ref{<dmm>} shows two dynamical maps, constructed in a dense grid of tens of thousands of initial conditions $(a_0,e_0)$. The red stars represent the initial condition $(a_0,e_0)=(\sim194,715 km,0.00267)$ of Methone at the initial date. The phase space in the vicinity of the orbit of Methone is characterized by three main regions:

a) The light region indicated by \textbf{C} around the current position of Methone in the maps is the domain of the Corotation resonance associated to the 15:14 Mimas-Methone mean-motion resonance. Defining $\psi\equiv15\lambda_{s}-14\lambda_M-\varpi_{M}$, where $s$ is a test satellite, we have that $\psi$ librates around zero for initial conditions in this whole region. The Corotation zone extends up to $e_0<0.015$ in the map for $194,655\leq a_0\leq 194,735$ km; at high eccentricities its domain is narrowed. In the map of the spectra of the semi-major axis, the Corotation zone suffers a slightly enlargement very close to the current elements of Methone. This particular aspect of our mappings of the Methone phase space shows good accordance with results of Rodr\'{i}guez and Callegari (2021) on the long-term stability of Methone (see figure 1 of that paper);

b) At the dark regions immediately close to the boundaries of the resonance in the dynamical maps, the spectra of the test satellites have large $N$ due to the proximity of the system to the boundaries of the 15:14 mean-motion resonance. In general, large values of $N$ can be associated to chaotic orbits in the separatrices of resonances, irregular or strong perturbed motion, as discussed before.

c) Regions of the maps where $N$ is small are in general associated to the loci in the phase space of regular motion linked to periodic orbits associated to resonances, and also to regions without physical significance in terms of resonant dynamics. In fact, all far regions in dynamical map with small $N$ are not associated to any resonance, as can be checked numerically with individual simulations.

%The position of Methone in DM of inclination is close to the border of the resonance, indicating a possible %complex long-term variation of the inclination of the clones of Methone. In fact, due to the forced inclination %in $i_{Me}\cos(\Delta\Omega_{Me-M})\times i_{Me}\sin(\Delta\Omega_{Me-M})$ plane (Fig. \ref{<secular>} in %Section 3.2) might lead to episodic oscillation of $\Delta\Omega_{Me-M})$ around zero at longer time spans in %numerical simulations\footnote{Aeageon is also affected by this kind of forcing term. See Fig. ????, Section %4.3.}. Moreover, the IPS shows the component \textbf{VLT} described above.

El Moutamid et al. (2014) studied the phase space of the Mimas-Methone resonance with an average two-degree of freedom model, ideal to be analyzed with the surface of section technique. In their figure 7, the y-axes of the sections are given by $\chi$, a suitable scaled action deduced within the domains of the model, representing the width of the resonance with the associated angle $\psi$\footnote{$\chi$ is defined in equation 25 in El Moutamid et al. (2014), and its expression can be obtained from their table 1. We utilize $\chi=k\left[\frac{ \sqrt{2} } {2} \left(\frac{a-a_0}{a_0}\right)  +14e^2\left(\frac{ \sqrt{2} } {2}-1\right)\right]$, where $a, e$ are the osculating semi-major axis and orbital eccentricity of the test particle; $k$ is a function of the masses, orbital elements of Mimas and $a_0$, the reference center of the Corotation zone, where we adopt $a_0=194,683$ km.}.

Thus, in order to give further analyses on the types of motion in the cases a) and b) listed above, we calculate numerically $\chi$ and plot it as a function of the corotation angle $\psi$ (e.g. Callegari and Yokoyama 2020). In Fig. \ref{<chi>}, we show several plots of a series of numerical orbits taken in the vicinity of Methone and in the regions of the separatrices. We have the following conclusions:

A) Inside the Corotation zone, where the orbit is regular, $\chi$ attains relatively small values $|\chi|\sim2$ or less, and depends on $\psi$ in a quasi-periodic manner such that $\psi$ librates. See red and magenta plots in Fig. \ref{<chi>} with initial conditions indicated by red star and magenta full disc in Fig. \ref{<dmm>}.
%These results are in good agreement with figure 7 in El Moutamid et al. (2014).

B) For chaotic orbits at the boundaries of the Corotation zone and eccentricities very close to zero, $\chi$ extends to slightly larger values while $\psi$ varies in irregular way such that $0\leq\psi\leq2\pi$ (in general, $\psi$ alternating between circulation and episodic oscillation in these cases). See green and pink plots in Fig. \ref{<chi>}.

Results A) and B) above can be checked by comparison with figure 7 in El Moutamid et al. (2014). They show good agreements on the determination of widths of the Corotation resonance and chaotic layer in spite of two distinct methods.

It is worth to stress the accuracy of the mapping of the phase space with dynamical maps: for instance, magenta and green initial conditions are separated by only 5 km in phase space (Fig. \ref{<dmm>}), and the results are confirmed: inside the Corotation zone, $\psi$ librates, which does not occur at the border\footnote{There is another additional indirect tests of our results: Fig. \ref{<chi>} was obtained from numerical simulations performed with the Mercury package (numerical scheme ii) described in Section 2), whereas in the calculation of the dynamical map, we applied our own code (scheme i))}.

C) For larger eccentricities in the chaotic zone, $|\chi|$ attains values much larger that 2, and its plot is diffused spread in $(\chi,\psi)$ plane (cyan and orange plots in Figs. \ref{<dmm>}, \ref{<chi>}). Orbits with such high $|\chi|$ are not given in El Moutamid et al. (2014), so that we have extended the mapping of chaotic orbits with $|\chi|$ not contemplated in El Moutamid et al. (2014). A similar result in the case of the small satellite Anthe is given in section 3.2.1 of Callegari and Yokoyama (2020).

\section{Conclusions}

Accurate characterization of the orbits of recently discovered Saturnian small satellites is a primary goal of planetary scientists since the beginning of the first results provided by Cassini spacecraft (e.g. Porco et al. 2005, Spitale et al. 2006). In this work, we provide an improvement in the interpretation of the resonant state of Methone (S/2004 S1) by numerically analyzing the main gravitational components on its orbit, which are mainly dictated by Mimas and the $J_2$ Saturn's field. We analyzed both the current orbit of Methone given in ephemeris platforms like the \emph{Horizons System}, as well numerical simulations with distinct numerical tools.

The 15:14 Mimas-Methone mean-motion resonance is well-characterized by the critical angle $\psi_1\equiv15\lambda_{Me}-14\lambda_M-\varpi_{M}$, which librates around zero with period $\sim521.57$ day and large amplitude $\sim 68$ degree. Physically, the libration of $\psi_1$  means that the conjunctions Mimas-Methone occur around a line which always oscillates in the direction of the pericenter of Mimas, corresponding to a typical characteristic of the so-called ``Corotation resonances''.

The libration of $\psi_1$ occurs independent of the type of elements we have adopted, namely, the geometric or osculating ones, since the mean-longitudes of Mimas and Methone, as well as the longitude of pericenter of Mimas, are not affected by the short-term induced variations due to $J_2$. When the geometric orbits are considered, it is observed oscillatory time variations of three other geometric critical arguments: $\psi_2=15\lambda_{Me}-14\lambda_M-\varpi_{Me}$, $\psi_3=15\lambda_{Me}-14\lambda_M+\varpi_{Me}-2\varpi_{M}$ and $\psi_4=15\lambda_{Me}-14\lambda_M+\varpi_{M}-2\varpi_{Me}$. This apparent conflicting resonant configuration can be explained by rearranging the expressions of the angular combinations such that they involve the relative variation of the longitudes of pericenters $\Delta\varpi_{Me-M}=\varpi_{Me}-\varpi_M$:
$\psi_2=\psi_1-\Delta\varpi_{Me-M}$, $\psi_3=\psi_1+\Delta\varpi_{Me-M}$, $\psi_4=\psi_1-2\Delta\varpi_{Me-M}$. Since the geometric  $\Delta\varpi_{Me-M}$ oscillates around zero with period $\sim2408.21$ day, the resulting time variations of $\psi_2$, $\psi_3$ and $\psi_4$ are compositions of the two frequencies, the resonant and the long-period ones (besides, evidently, all other perturbations of higher order and smaller magnitudes).

The above conclusions are immediate applications of classical results of resonant and secular theory (e.g. Brouwer and Clemence 1966, Murray and Dermott 1999) which, in the specific case of close-in satellites, a more general approach including the $J_2$ perturbations is necessary. Callegari and Yokoyama (2020) applied the generalized secular theory for the interpretation of the long-term variations in the orbit of the small satellite Anthe. Here we do the same in the case of Methone.
%Since the geometric $\omega_{Me}$ is free from fast circulation, the time variation of the geometric $\Delta\varpi_{Me-M}$ shows the long-term component $P_{\Delta\omega}$.

%In fact, it is this frequency, with period $P_{\psi_1}\sim521.57$ day which appears linked to the resonance in the plots of the orbital elements and corresponding spectra, as discussed in Section 3 (see Figs. \ref{<fft>}, \ref{<meth>}, Table 1).

The long-term oscillation of $\Delta\varpi_{Me-M}$ must only be guaranteed when the long-term time variations of the geometric longitudes of pericenter of Methone and Mimas are similar. In fact, spectral analyses of these angles show that their main components are centered at $\sim360.54$ day. Since the orbits of Mimas and Methone are separated by a few thousands of km, such correspondence of the main components cannot be explained only by the $J_2$ secular variations, which are responsible by the differential precessions of the orbits and are highly dependent on the distance from Saturn. The period of circulation of the longitude of pericenter of a satellite like Methone, owing to exclusively the $J_2$ perturbation, would be $\sim424.024$ day. The difference of this value to the current one is $\sim63.5$ day, and we have shown numerically that it is due to the 15:14 Corotation resonance. Thus, the resonance perturbations plays a fundamental role for the alignment of the orbits of Methone and Mimas.

We have mapped the domains of the 15:14 Mimas-Methone mean-motion resonance in the phase space. Unidirectional mappings give the values of the fundamental frequencies as functions of the initial osculating semi-major axis in the close vicinity of the current Methone orbit (Fig. \ref{<ipsm>}). The most appropriated mapping is the dynamical map defined in the plane $(a_0,e_0)$, where $e_0$ is the initial osculating eccentricity (Fig. \ref{<dmm>}). $\psi\equiv15\lambda_{s}-14\lambda_M-\varpi_{M}$, where $s$ is a test satellite, librates around zero in the maximum interval $194,655\leq a_0\leq 194,735$ km for $e_0<0.015$, delimiting the Corotation zone associated to the 15:14 Mimas-Methone mean-motion resonance. The boundaries of the Corotation zone are characterized by non-regular orbits, a typical motion associated to the separatrizes of the resonance. This result shows a good agreement with previous ones (e.g. El Moutamid et al. 2014). The orbit of Methone is located deeply inside the stable regions of the Corotation resonance, showing that it is regular.

\begin{acknowledgements}
We are grateful to Fapesp (Sao Paulo state research funding agency), through the processes 2019/15162-2, 2020/06807-7. We are especially grateful to two anonymous reviewers, who have made a detailed review of our work. We also thanks T. Yokoyama and M. T. dos Santos for discussions. T. Yokoyama is the main author of the program code utilized in this work to compute the geometric elements, which has been developed during the project of the work Callegari and Yokoyama (2020).
\end{acknowledgements}




\section*{Appendix 1: Initial conditions and parameters}
%\newpage

Table 3 gives the initial osculating elements and masses of the mid-sized satellites of Saturn and Methone provided by \emph{Horizons} system of ephemerides at date January 01, 2016.

The physical data for Saturn are $M_S=5.6834\times10^{26}$ (kg) (mass), $R_S=60,268\pm4$ km (equatorial radius), $J_2=0.01629071$ and $J_4=-0.0009358$ (Jacobson et al. 2006b).
\begin{table*}[h]
 \centering
\caption{Orbital elements and masses of Saturnian satellites taken from the \emph{Horizons} system refereed to date 2016-January-01, 00:00. Values collected from the update August 08, 2019. For Mimas and Methone, we also show the geometric elements calculated as described in Section 2. Assumed significant digits are purely formal. $\omega$, $\Omega$ are the argument of the pericenter and the longitude of ascending node, respectively.}
\vspace{0.5cm}
%\hline
%\hline
 \begin{tabular}{ccccc}
Satellite& semi-major axis  (km)            &  eccentricity                & inclination (degree)       &    mass (kg) \\
         &  $\omega$ (degree)                &    $\Omega$ (degree)         & mean anomaly (degree)               &                     \\
\hline
\hline
Mimas         &  186,025.6406619370&0.02002940475619249&1.568178165736470 &      3.75    \\
(osculating)  &  149.7425790886224 &92.66538879646959  &72.92980604437224 &              \\

Mimas         &  185,547.51453016 0&0.01951689931353337&1.56488701023253   &          \\
(geometric)   &  142.620036727989  &92.5697135272668      &80.1385659447950 &         \\

\hline

Methone       &  194,715.6734357022&0.02674177593041489&0.01386140166959750&$m^{(*)}$
 \\
(osculating)  &  247.6063949049330 &328.9309758484545& 356.9142844320854  &     \\

Methone       &  194,261.567967327 &0.0003677474052495034&0.01385532467227462&\\
(geometric)   &   267.725515393365 &328.878101566152    & 336.848107688711       &     \\

\hline

Enceladus    &     238,410.9497573420&0.005409700624203072 & 0.06300678158019136&10.805 \\
             &  254.6089765023341&107.0935554299123&54.19418982449177   &      \\
\hline
Tethys       &   294,975.1415797813 & 0.001057322770416529 & 1.093415017063881& 61.76  \\
             & 155.4378178629212 & 183.7764009032868&353.8714568422450  &\\
\hline
Dione        &   377,651.4603631170 &0.001723860642064985 &0.02979448678882671  &109.572\\
             &  161.8670464404854  &171.2052945850907 & 228.2313154582474 &\\
\hline
Rhea        &   5.272137290588300&0.01012735341245330 &0.3557125312486978  &230.9  \\
            &  18.62640796053094& 194.5283445769434 &27.80026501495894&    \\
\hline
Titan       &   1221,961.119150759  &0.02868590608459553&0.4023490211172993 & 13455.3\\
            &324.3272061447084  &247.8889323568444 &323.9021813775230              &\\
\hline
\end{tabular}

$(^{*})$ Not provided in \emph{Horizons} system. $m\sim4.21875\times10^{-7}$ $m_M$, where $m_M$ is the mass of Mimas, is an estimate of the value of the mass of Methone considering it with the same density of Mimas; we use it in the simulations of the numerical scheme i).
\end{table*}

\section*{Appendix 2: Critical arguments $\psi_k$, $5\leq k\leq14$}

\begin{figure*}
\centering
\includegraphics[width=12cm]{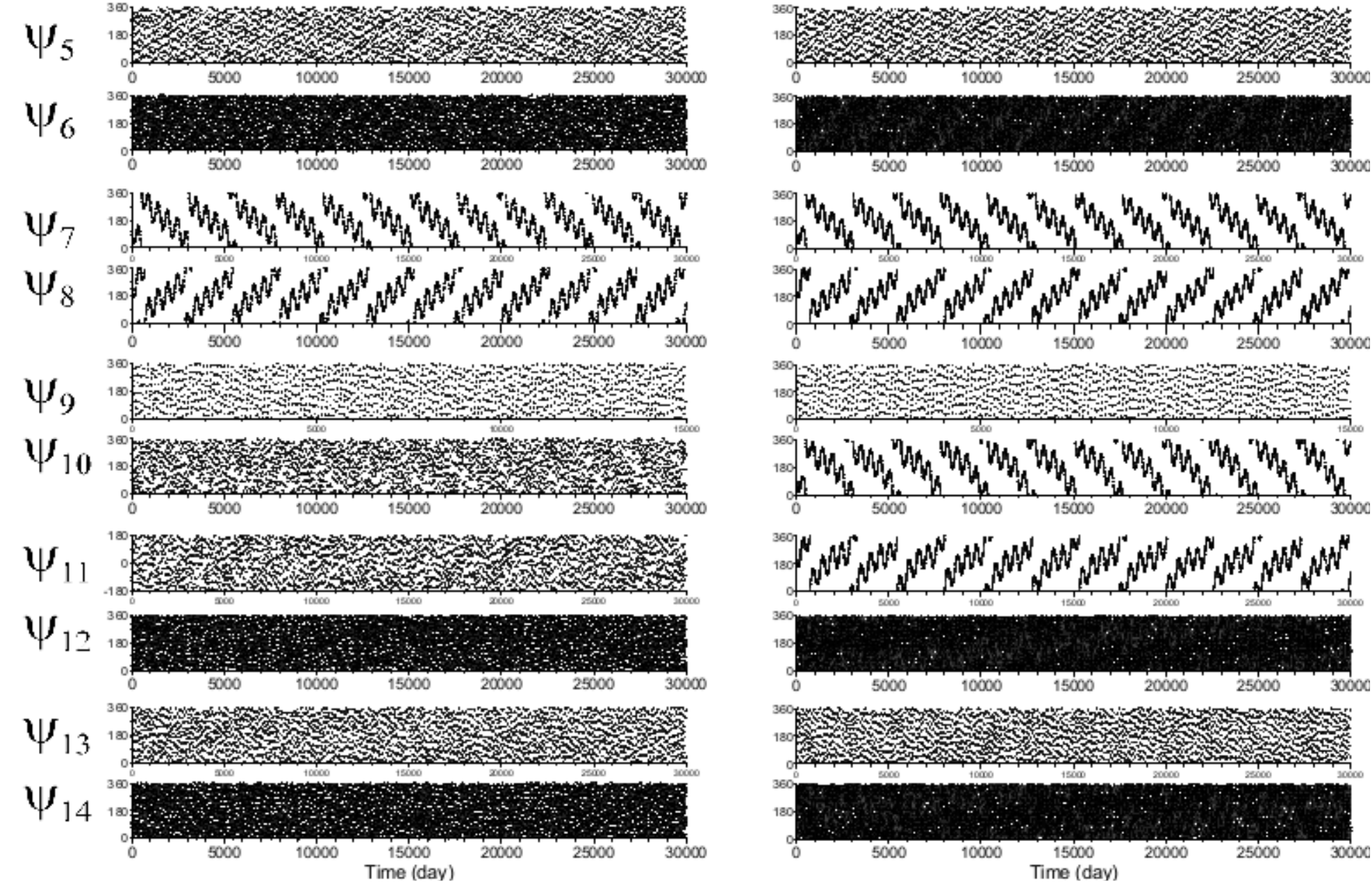}
     \caption{Fig. \ref{<meth>}, continued. $\psi_k$, $5\leq k\leq14$ are defined in Table 2.}
     \label{<meth2>}
\end{figure*}

In this Appendix, we extend the interpretation of the critical arguments of the disturbing given in Sections 3.1 in the case $\psi_k$, $k\geq5$ (Fig. \ref{<meth2>}).

The plots of the critical arguments of the disturbing $\psi_k$, $k\geq5$, are given in Fig. \ref{<meth2>}. All of these angles but $\psi_7$, $\psi_8$, $\psi_{10}$, $\psi_{11}$ do not have any physical relevance relevance since their time variations are dominated by short-term perturbations in both osculating and geometric variables.

From Table 2, we have:
\begin{eqnarray}
\psi_7&=&\alpha-\varpi_M-\Omega_{Me}+\Omega_M=\psi_1-\Delta\Omega_{Me-M},\label{3988}\\
\psi_8&=&\alpha-\varpi_M+\Omega_{Me}-\Omega_M=\psi_1+\Delta\Omega_{Me-M},\label{4888}
\end{eqnarray}
where $\alpha=15\lambda_{Me}-14\lambda_M$. $\psi_7$, $\psi_8$ are driven by the coupling of the oscillation of the corotation angle $\psi_1=\alpha-\varpi_M$ with a slower long-term circulation of $\Delta\Omega_{Me-M}$. Both, osculating and geometric $\Delta\Omega_{Me-M}$ circulate in prograde direction with a period of $P_{\Delta\Omega}\sim 2443.28$ day (recall from Section 3.2 that $\Delta\Omega_{Me-M}$ is not significantly affected by the induced short-term circulations due to $J_2$). The different signs of $\Delta\Omega_{Me-M}$ in (\ref{3988}), (\ref{4888}) cause the long-term circulation of $\psi_7$, $\psi_8$ to occur in the opposite directions.

From Table 2, we have:
\begin{eqnarray}
\psi_{10}&=&\psi_1-\Delta\varpi_{Me-M}-\Delta\Omega_{Me-M},\label{39881}\\
\psi_{11}&=&\psi_1-\Delta\varpi_{Me-M}+\Delta\Omega_{Me-M}.\label{48881}
\end{eqnarray}
Since the geometric $\Delta\varpi_{Me-M}$ oscillates around zero, the evolutions of the geometric $\psi_{10}$ and $\psi_{7}$, and $\psi_{11}$ and $\psi_{8}$, are respectively very similar.

\end{document}